\documentclass[12pt]{article}
\usepackage{amsfonts}
\usepackage{amssymb}
\usepackage{graphics,psboxit,amsmath}
\usepackage{subfigure}
\usepackage{graphicx}
\usepackage{verbatim}

\def\hybrid{\topmargin -30pt    \oddsidemargin 0pt 
        \headheight 0pt \headsep 0pt
        \textwidth 6.25in       
        \textheight 9.5in       
        \marginparwidth .875in
        \parskip 5pt plus 1pt   \jot = 1.5ex}

\hybrid

\def\baselinestretch{1.2}

\catcode`\@=11

\def\marginnote#1{}
\newcount\hour
\newcount\minute
\newtoks\amorpm
\hour=\time\divide\hour by60
\minute=\time{\multiply\hour by60 \global\advance\minute by-\hour}
\edef\standardtime{{\ifnum\hour<12 \global\amorpm={am}%
        \else\global\amorpm={pm}\advance\hour by-12 \fi
        \ifnum\hour=0 \hour=12 \fi
        \number\hour:\ifnum\minute<10 0\fi\number\minute\the\amorpm}}
\edef\militarytime{\number\hour:\ifnum\minute<10 0\fi\number\minute}

\def\draftlabel#1{{\@bsphack\if@filesw {\let\thepage\relax
   \xdef\@gtempa{\write\@auxout{\string
      \newlabel{#1}{{\@currentlabel}{\thepage}}}}}\@gtempa
   \if@nobreak \ifvmode\nobreak\fi\fi\fi\@esphack}
        \gdef\@eqnlabel{#1}}
\def\@eqnlabel{}
\def\@vacuum{}
\def\draftmarginnote#1{\marginpar{\raggedright\scriptsize\tt#1}}

\def\draft{\oddsidemargin -.5truein
        \def\@oddfoot{\sl preliminary draft \hfil
        \rm\thepage\hfil\sl\today\quad\militarytime}
        \let\@evenfoot\@oddfoot \overfullrule 3pt
        \let\label=\draftlabel
        \let\marginnote=\draftmarginnote
   \def\@eqnnum{(\theequation)\rlap{\kern\marginparsep\tt\@eqnlabel}%
\global\let\@eqnlabel\@vacuum}  }

\def\draft2{
        \def\@oddfoot{\sl preliminary draft \hfil
        \rm\thepage\hfil\sl\today\quad\militarytime}
        \let\@evenfoot\@oddfoot \overfullrule 3pt
        \let\label=\draftlabel
        \let\marginnote=\draftmarginnote
   \def\@eqnnum{(\theequation)\rlap{\kern\marginparsep\tt\@eqnlabel}%
\global\let\@eqnlabel\@vacuum}  }

\def\preprint{\twocolumn\sloppy\flushbottom\parindent 2em
        \leftmargini 2em\leftmarginv .5em\leftmarginvi .5em
        \oddsidemargin -.5in    \evensidemargin -.5in
        \columnsep .4in \footheight 0pt
        \textwidth 10.in        \topmargin  -.4in
        \headheight 12pt \topskip .4in
        \textheight 6.9in \footskip 0pt
        \def\@oddhead{\thepage\hfil\addtocounter{page}{1}\thepage}
        \let\@evenhead\@oddhead \def\@oddfoot{} \def\@evenfoot{} }

\def\numberbysection{\@addtoreset{equation}{section}
        \def\theequation{\thesection.\arabic{equation}}}

\def\underline#1{\relax\ifmmode\@@underline#1\else
        $\@@underline{\hbox{#1}}$\relax\fi}

\def\titlepage{\@restonecolfalse\if@twocolumn\@restonecoltrue\onecolumn
     \else \newpage \fi \thispagestyle{empty}\c@page\z@
        \def\thefootnote{\fnsymbol{footnote}} }

\def\endtitlepage{\if@restonecol\twocolumn \else \newpage \fi
        \def\thefootnote{\arabic{footnote}}
        \setcounter{footnote}{0}}  

\catcode`@=12
\relax

\def\figcap{\section*{Figure Captions\markboth
        {FIGURECAPTIONS}{FIGURECAPTIONS}}\list
        {Figure \arabic{enumi}:\hfill}{\settowidth\labelwidth{Figure
999:}
        \leftmargin\labelwidth
        \advance\leftmargin\labelsep\usecounter{enumi}}}
 \relax
\def\tablecap{\section*{Table Captions\markboth
        {TABLECAPTIONS}{TABLECAPTIONS}}\list
        {Table \arabic{enumi}:\hfill}{\settowidth\labelwidth{Table
999:}
        \leftmargin\labelwidth
        \advance\leftmargin\labelsep\usecounter{enumi}}}
 \relax
\def\reflist{\section*{References\markboth
        {REFLIST}{REFLIST}}\list
        {[\arabic{enumi}]\hfill}{\settowidth\labelwidth{[999]}
        \leftmargin\labelwidth
        \advance\leftmargin\labelsep\usecounter{enumi}}}
 \relax
\makeatletter
\newcounter{pubctr}
\def\publist{\@ifnextchar[{\@publist}{\@@publist}}
\def\@publist[#1]{\list
        {[\arabic{pubctr}]\hfill}{\settowidth\labelwidth{[999]}
        \leftmargin\labelwidth
        \advance\leftmargin\labelsep
        \@nmbrlisttrue\def\@listctr{pubctr}
        \setcounter{pubctr}{#1}\addtocounter{pubctr}{-1}}}
\def\@@publist{\list
        {[\arabic{pubctr}]\hfill}{\settowidth\labelwidth{[999]}
        \leftmargin\labelwidth
        \advance\leftmargin\labelsep
        \@nmbrlisttrue\def\@listctr{pubctr}}}
 \relax
\makeatother

\def\be{\begin{equation}}
\def\ee{\end{equation}}
\def\ba{\begin{eqnarray}}
\def\ea{\end{eqnarray}}

\def\del{\partial}

\def\k{\kappa}

\def\a{\alpha}

\def\b{\beta}

\def\G{\Gamma}
\def\d{\delta}
\def\D{\Delta}

\def\P{\Pi}

\def\th{\theta}
\def\Th{\Theta}
\def\m{\mu}

\def\om{\omega}
\def\Om{\Omega}

\def\L{\Lambda}
\def\s{\sigma}
\def\S{\Sigma}

\def\cN{{\cal N}}

\def\elF{{\bf F}}

\def\elPi{{\bf \Pi}}

\def\no{\noindent}

\def\qq{\qquad}

\def\IR{\relax{\rm I\kern-.18em R}}

\def \ov {\over}

\def\const{{\rm const.}}

\begin{document}


\renewcommand{\theequation}{\thesection.\arabic{equation}}
\csname @addtoreset\endcsname{equation}{section}

\newcommand{\eqn}[1]{(\ref{#1})}
\begin{titlepage}
\begin{center}

\hfill 0710.3162 [hep-th]\\

\vskip .5in

{\Large \bf String junctions in curved backgrounds,\\ their stability and
dyon interactions in SYM}

\vskip 0.5in

{\bf Konstadinos Sfetsos}\phantom{x} and\phantom{x} {\bf Konstadinos Siampos}
\vskip 0.1in

Department of Engineering Sciences, University of Patras,\\
26110 Patras, Greece\\

\vskip .1in

\vskip .15in

{\footnotesize {\tt sfetsos@upatras.gr},
\ \ {\tt ksiampos@upatras.gr}}\\

\end{center}

\vskip .4in

\centerline{\bf Abstract}

\no
We provide a systematic construction of string junctions in curved backgrounds
which are relevant
in computing, within the gauge/gravity correspondence,
the interaction energy of heavy dyons, notably of quark-monopole pairs,
in strongly coupled SYM theories.
To isolate the configurations of physical interest
we examine their stability under small fluctuations and prove several
general statements. We present all details, in several examples, involving non-extremal
and multicenter D3-brane backgrounds as well as the Rindler space.
We show that a string junction could be perturbatively stable even in branches that
are not
energetically the most favorable ones. We present a mechanical analog of this phenomenon.

\vfill
\no


\end{titlepage}
\vfill
\eject


\tableofcontents

\def\baselinestretch{1.2}
\baselineskip 20 pt
\no

\section{Introduction}

String junctions \cite{AhaSoYa,Schwarz} are of particular interest in string theory
for several reasons: Due to their BPS nature they can be
used to build  supersymmetric string networks \cite{Dasgupta,Sen,ReyYee,KroghLee, Matsuo} and
they are useful for
the description of BPS states in SYM theories \cite{Bergman} and for applications to gauge symmetry
enhancement in string theory \cite{Gabe}. The
scattering of string modes at string junctions in flat spacetime has been analyzed in \cite{CaTho}.
Particularly, important for the purposes in the present paper is the fact that
since string junctions connect strings of different type
they can be used, within the gauge/gravity correspondence \cite{adscft},
to compute the interaction energy of dyons
and in particular of heavy quark-monopole pairs \cite{Minahan}.

\no
The purpose of the present paper is to formulate and construct the string
junctions appropriate for computing the interaction potentials of heavy dyons
at strong coupling, within the
gauge/gravity correspondence. This is done for a large class of curved backgrounds
away from the conformal point and with reduced or no supersymmetry at all.
Having constructed string junction solutions does not imply that the corresponding dyon
interaction potentials are physical. One has to at least investigate
perturbative stability which will appropriately restrict the parametric space
in which the solutions are physically relevant.
Such investigations were exhaustively performed \cite{ASS1,ASS2}
for single string solutions useful in computing the heavy quark-antiquark potential.
Given these works and in comparison to them we will find
that the stability analysis of junctions leads to expected as well as
unexpected results, the latter being
counterintuitive to stability arguments based on energy considerations.

\no
The organization of this paper is as follows: In section 2 we formulate string junctions in a quite
general class of curved backgrounds. We derive general formulae from which the interaction
energy of a dyon pair is computed. In section 3 we consider in great detail the stability analysis
under small fluctuations of these string junctions aiming at discovering the physically relevant
regions. We pay particular attention to the precise formulation of the boundary and the
matching conditions at the junction point and derive general statements
that isolate the boundaries of
stability for all types of perturbative fluctuations.
In section 4, we present several examples of string junctions using D3-black branes and
multicenter D3-brane solutions which within the AdS/CFT correspondence are relevant for $\cN=4$ SYM at
finite temperature and at generic points in the Coulomb branch of the theory, respectively.
In addition, we present the details of string junctions in Rindler space, given its relevance
in a variety of black hole backgrounds near the horizon.
In section 5, we apply the outcome of the general stability analysis of section 3 to
the examples
of section 4. We present our conclusions in section 5. In the appendix  work out the stability of
a classical mechanical system which also exhibits an unexpected perturbative stability similar to the one
we found with string junctions.

\section{The classical solutions}
\label{sec-2}

In this section we develop and present the general setup of the AdS/CFT calculation of the static
potential of a heavy static dyon-dyon pair with general NS and RR charges.
These computations involve three strings and therefore we will be based
to the well-known results for the potential of heavy quark-antiquark pairs of \cite{maldaloop}
for the conformal case, as extended in \cite{bs} for general backgrounds,
and to the computation of the interaction energy of a heavy
quark-monopole pair in the conformal case \cite{Minahan}.

\no
We consider a general diagonal metric of Lorentzian signature of the form
\be
\label{2-1} ds^2 = G_{tt} dt^2 + G_{yy} dy^2 + G_{uu} du^2 +
G_{xx} dx^2 + G_{\th\th} d\th^2 + \ldots\ .
\ee
Here, $y$ denotes the (cyclic) coordinate along which the spatial
side of the Wilson loop extends, $u$ denotes the radial direction
playing the r\^ole of an energy scale in the dual gauge theory and
extending from the UV at $u \to \infty$ down to the IR at some
minimum value $u_{\rm min}$ determined by the geometry, $x$ stands
for a generic cyclic coordinate, $\th$ stands for a generic
coordinate on which the metric components may depend on and the omitted terms
involve coordinates that fall into one of the two latter classes without mixing terms.
It is convenient to introduce the functions
\ba
\label{2-2}
g(u,\th) &=& - G_{tt} G_{uu}\ ,\qq f_y(u,\th) = - G_{tt} G_{yy}\ ,
\qq f_x(u,\th) = - G_{tt} G_{xx}\ ,
\nonumber \\
\qq f_\th(u,\th) &=& - G_{tt} G_{\th\th}\ , \qq h(u,\th) = G_{yy}
G_{uu}\ . \ea If the conformal limit can be taken (this is the case
in all examples in the present paper except the one involving the
Rindler space) we can approximate the metric by that for $AdS_5\times
S^5$ with radii (in string units) $R=(4\pi g_s N)^{1/4}$, with $g_s$
being the string coupling. In this limit we have the leading order
expressions \be g \simeq h \simeq 1\ ,\quad f_x \simeq f_y \simeq
u^4/R^4\ ,\quad f_\th \simeq u^2 \ , \quad {\rm as} \quad u\to
\infty\ . \label{gh1} \ee

\no
A string junction consists of three co-planar strings joined at a point as depicted in Fig. 1.
The string labelled as 1
has $(p,q)$ NS and RR charges,
the second labelled as 2 has $(p',q')$ charges, whereas the third straight string labelled by 3 has
$(m,n)=(p+p',q+q')$ due to the charge conservation at the junction point.
From a microscopic point of view the strings that interact are the fundamental
$F$- and $D$-string with charges $(1,0)$ and $(0,1)$, respectively. Any other charge
combination should be achievable by performing in this pair an
$SL(2,\mathbb{R})$ transformation which gives the condition $p\ q^{\prime}-q\ p^{\prime}=\pm 1$ \cite{Gabe}.
The first two strings end at
the boundary of the space-time at $u=\infty $,
whereas the straight third string ends at the minimum value $u_{\rm min}$
of the space-time allowed by the geometry.
All three strings meet at the junction point at $u=u_0$.
Within the framework of the AdS/CFT correspondence, the interaction
potential energy of the $(p,q)$ dyon is given by
\be
\label{2-4}
e^{-{\rm i} E T} = \langle W(C) \rangle = e^{{\rm i} S_{p,q}[C]}\ ,
\ee
where
\be
\label{2-5}
S_{p,q}[C] = - {T_{p,q} \ov 2 \pi} \int d \tau d \sigma \sqrt{- \det g_{\a \b} }\ ,
\qq g_{\a\b} = G_{\mu\nu}
\partial_\alpha x^\mu \partial_\b x^\nu \ ,
\ee
is the Nambu--Goto action for a string propagating in the dual
supergravity background whose endpoints trace a rectangular contour
along one temporal and one space direction,
$T_{p,q}=\sqrt{p^2+q^2/g_s^2}$ and similar expressions hold for the strings 2 and 3.
We will not consider a contribution from the Wess--Zumino terms either because
these terms do not exist all or have vanishing contribution within our ansatz for solutions
below.

\no
We next fix reparametrization invariance for each string by choosing
\be
\label{2-6}
t=\tau \ ,\qq u=\s \ ,
\ee
we assume translational invariance along $t$, and we consider the
embedding
\be
\label{2-7}
y = y(u)\ ,\qq x =0\ ,\qq \th = \th_0 = \const\ ,\qq \hbox{rest} = \const\ ,
\ee
supplemented by the boundary condition
\be
\label{2-8}
u \left(-L_1\right)=u \left(L_2\right) = \infty\ ,
\ee
appropriate for a $(p,q)$ dyon placed at $y=-L_1$ and a $(p^\prime,q^\prime)$ dyon placed
at $y=L_2$.  The string with charges $(m,n)$ extending from the junction point
to $u=u_{\rm min}$ is
straight as $y=\const$ is always a solution of the equations of motion (see
\eqn{2-11} below).
\begin{figure}[!t]
\begin{center}
\begin{tabular}{cc}
\includegraphics[height=6cm]{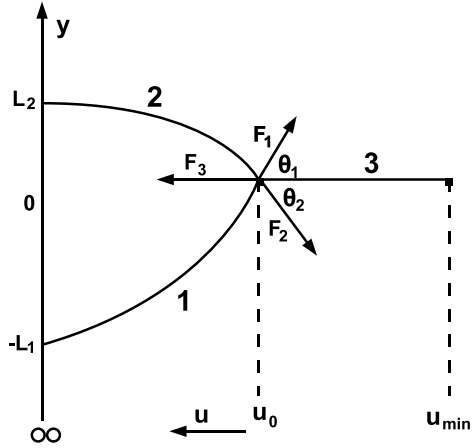}
\end{tabular}
\end{center}
\vskip -.5 cm \caption{Junction configuration: We denote by ${\bf F}_{i}$, $i=1,2,3$ (
${\bf F}_{pq}$ etc in the text)
the forces exerted at each string at the junction point $u_0$.
The turning points $u_i$, $i=1,2$
for the strings 1 and 2, are not depicted and lie at the right of $u_0$.
We always have $u_{\rm min}<u_0$,
but the $u_i$'s may or may not be larger than $u_{\rm min}$.}
\label{junction}
\end{figure}
In the ansatz \eqn{2-7}, the constant value $\th_0$ of the
non-cyclic coordinate $\th$ must be consistent with the
corresponding equation of motion. This requires that \cite{ASS1}
\be \label{2-9}
\partial_\th g(u,\th) |_{\th=\th_0} = \partial_\th f_y(u,\th) |_{\th=\th_0} = 0\ .
\ee
For the ansatz given above, the Nambu--Goto action reads
\be
\label{2-10}
S_{p,q} = - {T_{p,q}{\cal T}  \ov 2 \pi} \int d u \sqrt{ g(u) + f_y(u) y^{\prime 2}}\ ,
\ee
where ${\cal T}$ denotes the temporal extent of the Wilson loop, the
prime denotes a derivative with respect to $u$ while $g(u) \equiv
g(u,\th_0)$ and $f_y(u) \equiv f_y(u,\th_0)$ are the functions in
\eqn{2-2} evaluated at the chosen constant value $\th_0$ of $\th$.
Similar actions are also considered for the $(p',q')$ string, as well as for the
straight $(m,n)$ string.
Independence of the Lagrangian density
from $y$ implies that the associated momentum is conserved,
leading to the equation
\be
\label{2-11} {f_y y_{\rm cl}^\prime \ov \sqrt{ g + f_y y_{\rm
cl}^{\prime 2}}} = \mp f_{yi}^{1/2}\qq \Longrightarrow \qq y_{\rm
cl}^\prime = \mp {\sqrt{f_{yi} F_i}\ov f_y}\ ,
\ee
where $u_i$ are the values of $u$ at the turning point for each string, $f_{yi}
\equiv f_y(u_i)\ ,$ $f_{y0}\equiv f_y(u_0)$ and $y_{\rm cl}$ is the classical solution
with the
two signs corresponding to the lower (string 1) and upper (string 2).
The symbol $F_i$
(not to be confused with the forces in Fig. \ref{junction}) stands for the function
\be
\label{2-14} F_i = {g f_y \ov f_y - f_{yi}}\ ,\qq i=1,2\ .
\ee
Note that, for the junction point $u_0$ we have that $u_0\geqslant u_{\rm min}$.
For the turning points we have that $u_i\leqslant u_0$,
but they are not necessarily larger than $u_{\rm min}$,
since the strings 1 and 2 are not actually
extending to their turning points.
Integrating \eqn{2-11}, we express the separation length as
\be
\label{2-12}
L =L_1+L_2=f_{y1}^{1/2}
\int_{u_0}^{\infty} d u {\sqrt{F_1} \ov f_y}+f_{y2}^{1/2} \int_{u_0}^{\infty}
d u {\sqrt{F_2} \ov f_y}\ .
\ee
Finally, inserting the solution for $y_{\rm cl}^\prime$ into
\eqn{2-10} and subtracting the divergent self-energy contribution
of disconnected worldsheets, we write the interaction energy of the junction as
\ba
\label{2-13}
&&E = {1 \ov 2\pi}\left(T_{p,q}{\cal E}_1  +T_{p^\prime,q^\prime} {\cal E}_2
+ T_{m,n} \int_{u_{\rm min}}^{u_0} d u \sqrt{g}\right)\ ,\nonumber \\
&&{\cal E}_i=\int_{u_0}^\infty d u \sqrt{F_i} -
\int_{u_{\rm min}}^\infty d u \sqrt{g}\ ,\qq i=1,2\ .
\ea
Ideally, one would like to evaluate the integrals \eqn{2-12}
and \eqn{2-13} exactly, solve \eqn{2-12} for $u_0$ and insert
into \eqn{2-13} to obtain an expression for the energy $E$ in
terms of the separation length $L$ and $u_1,u_2$.
Then, minimizing this expression with respect to
the $u_i$'s we will obtain the expression of the
$E$ in terms of the separation length $L$, which is then identified with the interaction
energy of the heavy dyon pair.

\no
However, in practice this
cannot be done exactly, except for the conformal case,
and Eqs. \eqn{2-12} and \eqn{2-13} are to be regarded as
parametric equations for $L$ and $E$ with parameters $u_0,u_1,u_2$.
A much easier way to find $u_1,u_2$ is to impose that the net force at the string junction is
zero \cite{Schwarz,Sen,CaTho}, an approach followed in the present context
for the conformal case in \cite{Minahan}.
The infinitesimal lengths squared along each string are
\be
d\ell_i^2=(G_{yy} y'^2_{\rm cl} + G_{uu}) du^2 = -{1\ov G_{tt}}F_i du^2\ ,\qq i=1,2\ .
\ee
Hence from the action
\eqn{2-10}
we find that the tensions of the strings at the junction point $u=u_0$ are
$ \displaystyle{T_{p,q}\ov 2\pi}\sqrt{-G_{tt}}$ and similarly for the other two string that meet at
the junction.
The angles between the string and the $u$ axis at the junction point are computed from
\ba
\label{2-14a}
{\rm At}\ u=u_0:\qq \tan \th_i = \Bigg| {\sqrt{G_{yy}} dy\ov \sqrt{G_{uu}} du}\Bigg |
\qq \Longrightarrow \qq
\sin\th_i=\sqrt{{f_{yi}\ov f_{y0}}}\ , \qq i=1,2\ ,
\ea
where we used \eqn{2-11} and took into account that displacements along perpendicular
axes are measured by a curved metric. The above explicitly
shows that the angles depend on the parameters $u_0$ and $u_i$.
The forces exerted to each of the strings on the $u\!-\!y$ plane are
\ba
&&{\bf F}_{p,q}={T_{p,q}\ov 2\pi}\sqrt{-G_{tt}}\ (-\cos\th_1,\sin\th_1)\ ,
\nonumber  \\
&&{\bf F}_{p^{\prime},q^{\prime}} = -{T_{p^{\prime},q^{\prime}}\ov 2\pi}\sqrt{-G_{tt}}\ (\cos\th_2,\sin\th_2)\ ,
\label{2-16}\\
&&{\bf F}_{m,n} = {T_{m,n}\ov 2\pi}\sqrt{-G_{tt}}\ (1,0)\ ,
\nonumber
\ea
the first entry being the $u$ and the second the $y$ component.
Demanding that the total force be zero one finds that the angles at the equilibrium point are given by
\ba
\label{2-14b}
\cos\th_1 &=& {T_{m,n}^2+T_{p,q}^2-T_{p^{\prime},q^{\prime}}^2\ov{2 T_{m,n}T_{p,q}}}\ , \nonumber\\
\cos\th_2 &=& {T_{m,n}^2+T_{p^{\prime},q^{\prime}}^2-T_{p,q}^2\ov{2 T_{m,n}T_{p^{\prime},q^{\prime}}}}\ .
\ea
From these expressions the angles are determined in terms of the NS and RR charges of the
strings and then from \eqn{2-14a} one may express $u_i$ in terms of $u_0$ and the string
charges.
Recalling that $u_0$ is determined by $L$, one ends with the interaction energy being solely a
function of $L$ and of the strings' charges as fixed parameters.

\no
We can verify that the approach of fixing the parameters $u_i$ by utilizing the zero force condition at the
junction point gives rise to a minimal energy. Since we can not solve \eqn{2-12} for $u_0$
in general, we shall recall that we have an implicit
expression of $u_0$ in terms of $L,u_1,u_2$, by inverting \eqn{2-12}. Since we choose to minimize
the energy with respect to $u_i$, the physical
length $L$ does not depend on them and we have
\ba
\label{2-15a}
{\partial L\ov\partial u_i}=0\ ,\qq i=1,2\ ,
\ea
\no
which can be used together with
\eqn{2-14a} to express the derivative of $u_0$ with respect to $u_i$ as
\ba
\label{2-15b}
{\partial u_0\ov\partial u_i}={1\ov \tan\th_1+\tan\th_2}\ {f_{yi}^\prime\ov 2}\ \sqrt{f_{y0} \ov g_0f_{yi}}\
\int_{u_0}^\infty du {\sqrt{gf_y}\ov(f_y-f_{yi})^{3/2}}\ ,\qq i=1,2\ .
\ea
We have not used the no force condition to derive the above in the sense that
the angles $\th_i$ are not constant except for the equilibrium point.
Next, using \eqn{2-15b} and the zero force condition one can check, after careful algebraic manipulations, that
\be
{\partial E\ov\partial u_i}=0\ ,\qq i=1,2 \ ,
\label{eui}
\ee
as advertized.
Our proof is completely general within our ansatz \eqn{2-1}
and no-where we did need
to explicitly evaluate the various integrals by resorting to
specific examples.\footnote{
The result is also valid even when in the upper limit of integration infinite is replaced by a
finite value, as in the example of the Rindler space below.}
Had we done so, in the cases where this is possible,
we would have ended up with complicated expressions and our proof would have been equivalent to proving
certain identities involving, in general, special functions (for instance,
for the conformal case see the app. of \cite{Minahan}).

\no
Let us mention that the
expression for the interaction energy $E$ in terms of length $L$ for  the configuration
of $(p,q)$ and $(q,p)$ dyons is invariant under the S--duality transformation $g_s\to 1/g_s$,
as it should be. However, this is not true for $L$ and $E$ separately as functions of the
auxiliary parameter $u_0$.

\subsection{On the quark-monopole interaction}

The most important case arises for the interaction potential of a quark with a monopole
with charges $(1,0)$ and $(0,1)$, respectively.
In that case it is easily seen from \eqn{2-14b} that
$\th_1+\th_2=\pi/2$. It will be useful for later purposes to
consider the limit of $g_s\to 0$. In that case the tension
of the $(0,1)$ string corresponding to the monopole becomes very large,
so that we expected that it stays
almost straight, whereas in order for the forces to
balance, the $(1,0)$ string should hit the straight string
almost perpendicularly. This is indeed reflected in the expansions
\ba
\th_1 & = & {\pi\ov 2} - g_s + {1\ov 3} g_s^3 + {\cal O}(g_s^5)\ ,
\nonumber\\
\th_2 & = &  g_s - {1\ov 3} g_s^3 + {\cal O}(g_s^5)\ .
\label{th12}
\ea
Of course an equivalent expansion exists for $g_s\to \infty$,
with $\th_1$ and $\th_2$ interchanged and $g_s\to 1/g_s$.
In the small string coupling limit we easily infer from \eqn{2-14a} and \eqn{th12}
that the turning points of the strings 1 and 2 are
\ba
u_1 = u_0 + {\cal O}(g_s^2)\ ,\qq u_2 = u_{\rm root} + {\cal O}(g_s^4)\ ,\qq
\label{u12}
\ea
where $u_{\rm root}$ is the largest root of the equation $f_{y2}=0$.
In the examples we present in detail below, $u_{\rm root}=u_{\rm min}$,
except for one case in which it has a smaller value.
Therefore the expressions for the separation length \eqn{2-12} and the energy \eqn{2-13}
for the quark-monopole interaction, become
\ba
\label{2-12li}
L =f_{y0}^{1/2}
\int_{u_0}^{\infty} d u {\sqrt{F_0} \ov f_{y}} + {\cal O}(g_s) \
\ea
and
\ba
\label{2-13li}
E = {1 \ov 2\pi} \left( \int_{u_0}^\infty d u \sqrt{F_0} -
\int_{u_{\rm min}}^\infty d u \sqrt{g}\right) + {\cal O}(g_s)\ ,
\ea
where we have defined $F_0$ as in \eqn{2-14} with $f_{yi}$ replaced by $f_{y0}=f_y(u_0)$.
Thanks to the $S$-duality invariance of
the general results \eqn{2-12} and \eqn{2-13} identical expressions to \eqn{2-12li} and
\eqn{2-13li} hold in the large string coupling expansion, where in the correction terms
we just replace $g_s\to 1/g_s$.
Hence, in the limit of small and large string coupling,
the energy and the length assume half of their corresponding values
for the single string for the quark-antiquary system.
We emphasize that this fact does note imply that
the spectrum of small fluctuations around the each classical configuration will be
the same. Actually we will explicitly demonstrate that the opposite is true, which
implies that no matter how stiff two of the strings (the straight and one of the strings
1 or 2) become, its fluctuations do not decouple and affect those of the third string.

\section{Stability analysis}
\label{sec-3}

We now turn to the stability analysis of
these configurations, aiming at isolating in parametric
space the physically stable regions and pointing out the similarities
and differences with the similar analysis for string
configurations dual to the quark-antiquark potential.
The small fluctuations about the classical
solutions we will discuss, fall into three types: (i) ``transverse''
fluctuations, referring to cyclic coordinates transverse to the
dyon-dyon axis such as $x$, (ii) ``longitudinal''
fluctuations, referring to the cyclic coordinate $y$ along the
dyon-dyon axis, and (iii) ``angular'' fluctuations,
referring to the special non-cyclic coordinate $\th$.
In \cite{ASS1} general results for single string fluctuations of all of the above types
for the class of backgrounds
with metric \eqn{2-1} were derived and used to analyze the stability of heavy quark-antiquark
potentials. These are certainly relevant for the investigations of the present paper, so that we
review them below, but one has to be
careful by paying particular attention to the matching conditions at the junction point of the three strings
which affect the parametric space where the stability occurs, in crucial ways.

\subsection{Small fluctuations}

We parametrize the fluctuations about the equilibrium configuration,
by perturbing the embedding according to
\be
\label{3-1}
x_i = \d x_i (t,u)\ ,\qq y_i = y_{\rm cl,i}(u) + \d y_i (t,u)\ ,\qq \th_i = \th_0 + \d \th_i(t,u)\ ,
\ee
where the index $i$ refers to the three strings forming the junction. Note that we have kept
the gauge choice \eqn{2-6} unperturbed by using worldsheet
reparametrization invariance. We then calculate the Nambu--Goto action for this
ansatz and we expand it in powers of the fluctuations.  The zeroth-order term gives
just the classical action and the first-order vanishes thanks to the classical
equations of motion.\footnote{A careful treatment of the first-order terms gives rise to boundary
terms. Demanding that they vanish turns out to be equivalent to the zero-force condition at the junction point.
This can be shown quite straightforward for junctions in a flat spacetime before a gauge choice is made
and reproduces the conditions found, for instance, in \cite{CaTho}.
However, in curved spaces after the gauge choice $u=\s$ is made, one obtains straightforwardly
only the $y$-component
of the zero-force condition. Switching to the gauge $y=\s$ one obtains the
$u$-component as well.} The
resulting expansion for the quadratic fluctuations for the $i=1,2$ string is written as \cite{ASS1}
\ba
\label{3-4} \!\!\!\!\!\!\!\!\!\!\!\! S^{(i)}_2 &=& - {1 \ov 2\pi} \int
dt du \biggl[ {f_x \ov 2 F_i^{1/2}}
\d x_i^{\prime 2} - {h f_x F_i^{1/2} \ov 2 g f_y} \d \dot{x_i}^2 \nonumber\\
 && \qq\qq\quad\; + {g f_y \ov 2 F_i^{3/2}} \d y_i^{\prime 2} - {h \ov 2 F_i^{1/2}} \d \dot{y_i}^2 \\
&& \qq\qq\quad\; + {f_\th \ov 2 F_i^{1/2}} \d \th_i^{\prime 2} - {h
f_\th F_i^{1/2} \ov 2 g f_y} \d \dot{\th_i}^2 + \left( {1 \ov 4
F_i^{1/2}}
\partial_\th^2 g
+ {f_{y0} F_i^{1/2} \ov 4 f_y^2} \partial_\th^2 f_y \right) \d \th_i^2
\biggr]\ ,
\nonumber
\ea
where we have used, the condition \eqn{2-9} and all functions and their $\th$--derivatives are again
evaluated at $\th=\th_0$. Writing down the Euler--Lagrange equations for this action,
 using independence of the various functions from $t$ and  setting
\be
\label{3-5}
\d x_i^\m (t,u) = \d x_i^\m (u) e^{-{i} \omega t}\ ,\qq i=1,2,3\ ,
\ee
we find that for the strings 1 and 2 the linearized equations for the three types
of fluctuations read
\ba
\label{3-6}
&&\left[ {d \ov du} \left({f_x \ov F_i^{1/2}} {d \ov du} \right)
+ \omega^2 {h f_x F_i^{1/2} \ov g f_y} \right] \d x_i = 0\ ,
\nonumber\\
&&\left[ {d \ov du} \left( {g f_y \ov F_i^{3/2}} {d \ov du} \right)
+ \omega^2 {h \ov F_i^{1/2}} \right] \d y_i = 0\ ,
\label{cunv}
\\
&&\left[ {d \ov du} \left({f_\th \ov F_i^{1/2}} {d \ov du} \right)
 + \left( \omega^2 {h f_\th F_i^{1/2} \ov g f_y}
- {1 \ov 2 F_i^{1/2}} \partial_\th^2 g - {f_{y0} F_i^{1/2} \ov 2
f_y^2} \partial_\th^2 f_y \right) \right] \d \th_i = 0\ . \nonumber
\ea
For the straight string the action for the quadratic fluctuations is
\ba
S^{(3)}_2 &= & -{1\ov 4\pi}
 \int dt du \biggl[{f_y\ov\sqrt{g}}\d y_3^{\prime 2}+{f_x\ov\sqrt{g}}\d x_3^{\prime 2}+{f_{\th}\ov\sqrt{g}}
\d \th_3^{\prime 2}-
{h\ov\sqrt{g}}\d\dot{y}_3^2-{f_x h\ov \sqrt{g}f_y}\d\dot{x}_3^2
\nonumber\\
&&\phantom{xxxxxxxxxx}
-{f_{\th} h\ov \sqrt{g}f_y}\d\dot{\th}_3^2  +{\del^2_\th g\ov 4 \sqrt{g}} \d\th_3^2\biggr]\ ,
\ea
from which we obtain the equations
\ba
&&
{d\ov du}\left({f_x\ov\sqrt{g}}\d x_3^{\prime}\right)+\omega^2{hf_x\ov\sqrt{g}f_y}\d x_3=0\ ,
\nonumber \\
&&{d\ov du}\left({f_y\ov\sqrt{g}}\d y_3^{\prime}\right)+\omega^2{h\ov\sqrt{g}}\d y_3=0\ ,
 \label{jklo}\\
&&{d\ov du}\left({f_{\th}\ov\sqrt{g}}\d \th_3^{\prime}\right)
+\omega^2{hf_{\th}\ov\sqrt{g}f_y}\d\th_3
={1\ov 2}{\partial_\th^2 g\ov\sqrt{g}}\d\th_3\ .
\nonumber
\ea
Note that for the straight string the equations can be obtained by
formally letting $F_i\to g$ and $f_{y0}\to 0$ in \eqn{cunv}.

\no
Hence determining the stability of the string
configurations of interest has been reduced to a
eigenvalue problem of the general Sturm--Liouville type
\be
\label{3-7} \left\{ - {d \ov du} \left[ p(u;u_0) {d \ov du}
\right] - r(u;u_0)
 \right\} \Phi(u) = \omega^2 q(u;u_0) \Phi(u)\ ,\quad u_{\rm min}\leqslant u_0\leqslant u
< \infty\ ,
\ee
where the functions $p(u;u_0)$, $q(u;u_0)$ and $r(u;u_0)$ are read
off from \eqn{3-6} and \eqn{jklo} and depend parametrically on $u_i$ through the
function $F_i$ in \eqn{2-14}, which in turn is determined by \eqn{2-14a} in terms of
the junction point $u_0$ and the NS and RR charges of the strings. Our aim is to find the
range of values of $u_0$ for which $\omega^2$ is negative.
In fact the extremal such value of $u_0$ (the minimum as it turns out)
will be decided by determining the zero mode $\om=0$, which is an
easier problem.
Note also that, although
it may seem so,
the equations above for the three strings and for given type of fluctuations are not decoupled.
The matching
conditions at the junction point couple them in an essential manner.
This equation can not be solved analytically in general, so that
sometimes it is convenient to transform the Sturm--Liouville into a Schr\"odinger equation
and use known approximating analytical methods that boil down to simple numerical problems.

\subsection{Boundary and matching conditions}

To fully specify our eigenvalue problem, we must impose
boundary conditions on the fluctuations at the UV
limit $u\to\infty$, at the IR limit $u=u_{\rm min}$, as well as matching conditions at the junction
point at $u=u_0$.
The boundary condition at the UV are chosen to be
\be
\Phi(u) = 0\ ,\qq {\rm as}\quad u\to \infty \ ,
\label{3-14b}
\ee
where $\Phi$ is any of the fluctuations for the strings 1 and 2. These
represent the fact that we keep the dyons at fixed points at the boundary.
In the far IR at $u=u_{\rm min}$ we simply demand, for the string 3 that extends also in there,
finiteness of the solution and its $u$-derivative.
Otherwise,
all perturbations might be driven away from small values.

\no
We demand that the $\d x$ fluctuations as well
as their variations in the
quadratic actions should be equal at the junction point $u=u_0$, so that the latter does not break.
Moreover, we demand that the total boundary term resulting in deriving the classical
equation of motions from the quadratic actions $S_2^{(i)}$ vanishes. These two requirements
give rise to
\ba
\label{3-18a}
&&\d x_1=\d x_2=\d x_3\ , \qq {\rm at} \quad u=u_0\ ,
\nonumber \\
&&T_{p,q}\cos\th_1\d x_1^{\prime}+T_{p^\prime,q^\prime}\cos\th_2\d x_2^{\prime}
-T_{m,n}\d x_3^\prime=0\ , \qq {\rm at} \quad u=u_0\ ,
\ea
where we have  used \eqn{2-14a} to simplify the second condition.

\no
Identical
conditions with those in \eqn{3-18a} hold for the angular fluctuations $\d\th$ as well
\ba
\label{3-18ab}
&&\d \th_1=\d \th_2=\d \th_3\ , \qq {\rm at} \quad u=u_0\ ,
\nonumber \\
&&T_{p,q}\cos\th_1\d \th_1^{\prime}+T_{p^\prime,q^\prime}\cos\th_2\d \th_2^{\prime}
-T_{m,n}\d \th_3^\prime=0\ , \qq {\rm at} \quad u=u_0\ .
\ea

\no
The appropriate boundary and matching conditions for the $\d y$
fluctuations should be found first in
a coordinate system in which the classical solution does not change.
This is simply given by \cite{ASS1}
\be
\label{3-17}
u=\bar u + \d u(t,u) \ ,\qq \d u(t,u) = - {\d y(t,u)\ov
y^\prime_{\rm cl}(u)}\ ,
\ee
and is easily checked that the classical solution is not perturbed at all. This redefinition
does not affect the $\d x$- and $\d \th$--fluctuations since they have
trivial classical support and we keep only linear, in fluctuations,
terms. At the junction point the $\d u$ fluctuations are continuous and this
leads to a discontinuity for the $\d y$ fluctuations due to \eqn{3-17}
for the strings 1 and 2.
As before, we demand that the total boundary term resulting in deriving the classical
equation of motions from the quadratic actions $S_2^{(i)}$ vanishes, keeping in mind that
the variations of the $\d y_i$'s in their quadratic actions should
be equal at the junction point.
With the use of \eqn{2-14a}
we have the following conditions for the $\d y$ fluctuations
\ba
&&\d y_1\cot\th_1+\d y_2\cot\th_2=0\ ,\qq {\rm at} \quad u=u_0\ ,
\nonumber \\
&&T_{p,q}\cos^3\th_1\d y_1^{\prime}+T_{p^\prime,q^\prime}\cos^3\th_2\d y_2^{\prime}
-T_{m,n}\d y_3^\prime=0\ ,\qq {\rm at} \quad u=u_0\ .
\label{3-11}
\ea

\subsection{Zero modes}

As we have noted we can obtain the boundary of stability in parametric space by
studying the zero-mode problem.  Hence, we examine closely the three distinct types of fluctuations.

\subsubsection{Transverse zero modes}

For the transverse fluctuations the zero mode
solution obeying \eqn{3-14b} for the strings 1 and 2 is
\ba
\d x_i   =  a_i I_i(u)\ ,
\qq I_i(u)=\int^{\infty}_u {du\ov f_x} \sqrt{gf_y\ov f_y-f_{yi}}\ ,\qq i=1,2\ ,
\ea
where the $a_i$'s are  multiplicative constants.
The zero mode solution for the straight string is
\ba
\d x_3=a_3\int^{u_{0}}_u du {\sqrt{g}\ov f_x}+\const
\ea
When we specialize this expression in our examples we will see that either $\d x_3$ and/or
$\d x_3^\prime$ diverges when $u\to u_{\rm min}$, so we choose
$a_3=0$ in order for the solution and its derivative to remain finite.
Thus $\d x_3=\const$ and the zero mode solution becomes irrelevant for the stability analysis.
From \eqn{3-18a} and the zero force condition,
one finds a linear homogeneous system of algebraic equations for $a_1$ and $a_2$, whose
determinant should vanish for non-zero solutions to exist. In this way we find the following condition
\ba
\sin\th_1 I_1(u_0)+\sin\th_2 I_2(u_0)=0\ .
\ea
\no
Since $I_i(u)>0$ this condition cannot be satisfied and therefore the transverse zero
modes do not exist, rendering the corresponding fluctuating modes as stable.
Our finding is similar to the conclusion that the transverse fluctuations of the single string
fluctuation corresponding to the potential of heavy quark-antiquark pairs are also stable \cite{ASS1}.

\subsubsection{Longitudinal zero modes}

We next turn to the longitudinal fluctuations. The zero mode
solution obeying \eqn{3-14b} for the strings 1 and 2 is
\ba
\d y_i   = b_i J_i(u)\ ,\qq  J_i(u)=\int^{\infty}_u du {\sqrt{gf_y}\ov (f_y-f_{yi})^{3/2}}\ ,\qq i=1,2 \ ,
\ea
where the $b_i$'s are multiplicative constants.
The zero mode solution for the straight string is
\ba
\d y_3=b_3\int^{u_0}_u du{\sqrt{g}\ov f_y}+ \const
\ea
As before, when we specialize this expression in our examples we will see that either
$\d y_3$ and/or $\d y_3^\prime$ diverges when $u\to u_{\rm min}$, so we choose
$b_3=0$ in order for the solution and its derivative to remain finite.
 Therefore the zero mode fluctuation becomes
irrelevant for the stability analysis.
Similarly to before, from \eqn{3-11} and the zero force condition one finds the following relation
\ba
\label{3-18}
\cos\th_1 J_1(u_0)=\cos\th_2 J_2(u_0)\ .
\ea
This is an equation for $u_0$ in terms of the string coupling and the strings' charges,
which can be solved numerically. We will denote its solution, whenever it exists,
by $u_{0  c}$.
In general we will have expansions of the form
\ba
u_{0c}(g_s)& = & u^{(0)}_{0c}+ \sum_{n=1}^\infty u_n^{(0)} g_s^n\ ,
\nonumber\\
u_{0c}(g_s)& = & u^{(\infty)}_{0c}+ \sum_{n=1}^\infty u^{(\infty)}_n g_s^{-n}\ .
\label{exppa}
\ea
In the most important case of a quark-monopole interaction, $S$-duality invariance of the
result dictates that $u_{0c}(g_s)=
u_{0c}(g_s^{-1})$ and therefore the two perturbative expansions
above contain the same information. Namely,
$ u^{(0)}_{0c}=  u^{(\infty)}_{0c}$ and all coefficients are equal,
i.e. $ u_n^{(0)}=u^{(\infty)}_n$.
Actually, then the critical value $u_{0c}$
does not change significantly from its leading order values $u^{(0)}_{0c}$, the reason
being that the small and large $g_s$ results actually coincide and therefore
there is not much room for big changes in between.
We show below that the leading term $u_{0c}^{(0)}$ can be computed by \eqn{limgs}
which is an approximative version of \eqn{3-18}.
In the examples below
the leading order result for $u_{0c}$ from \eqn{limgs} and that computed for $g_s\simeq 1$ using \eqn{3-18}
differ by two to ten percent.

\begin{figure}[!t]
\begin{center}
\begin{tabular}{cc}
\includegraphics[height=5.2cm]{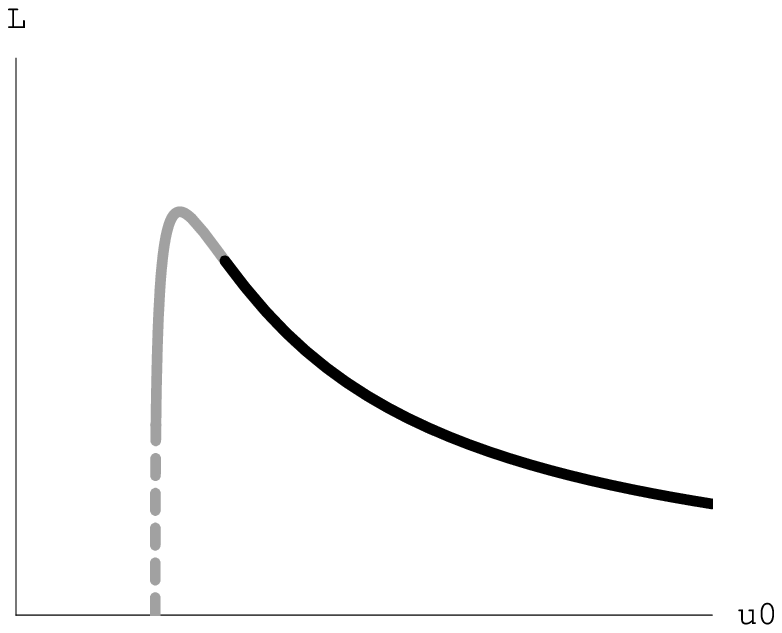}
&\includegraphics[height=5.2cm]{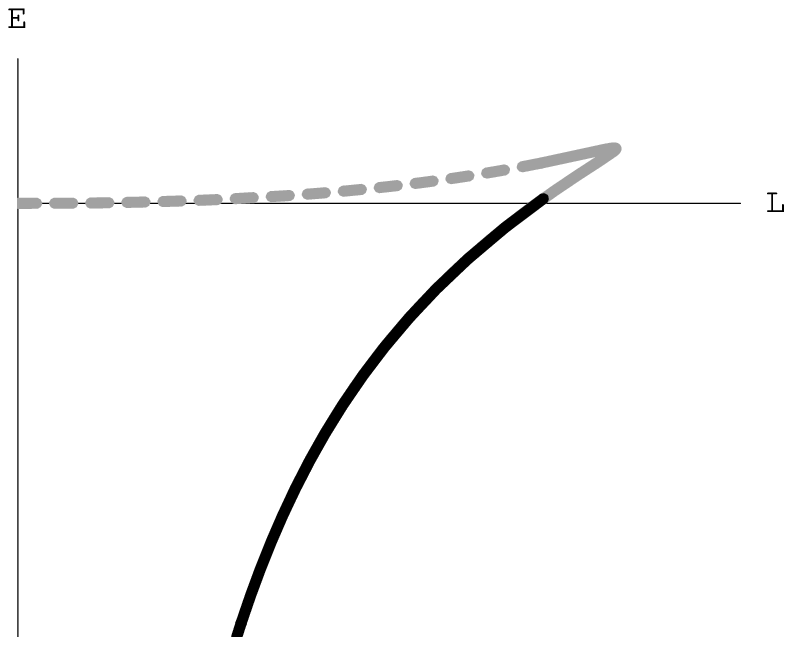}\\
(a) & (b)
\end{tabular}
\end{center}
\vskip -.5 cm \caption{Plots of $L(u_0)$
 and $E(L)$ for cases where there is an instability of the longitudinal modes.
This covers the case of non-extremal D3-brane, multicenter $D3$-branes on
a sphere for the $\th_0=0$ trajectory and the Rindler space.
The various types of lines correspond to stable (solid dark),
 metastable (solid gray) and unstable (dashed gray) configurations,
 with the stability determined by the analysis of section 5.}
\label{fig1}
\end{figure}

\no
A natural question is whether or not the value of $u_{0c}$ coincides with that corresponding
to the maximum value of $L(u_0)$, labelled by $u_{0m}$ if that exists,
below which the energetically less favorable
branch in the $E\!-\! L$ diagram starts (a typical case is presented in fig. \ref{fig1}).
This might have been expected since in \cite{ASS1,ASS2}
it was shown quite generally that an instability of longitudinal fluctuations of a single string
used to compute the heavy quark-antiquark potential,
occurs precisely at the solutions of the equation $L'(u_0)=0$ (which is identical to $E'(u_0)=0$).
In our case
following similar manipulations as in \cite{ASS1} and using \eqn{2-14a} at the
equilibrium point where the $\th_i$'s are constant, in order to find
the derivatives of $u_i$ with respect to $u_0$, we obtain
\ba
\label{3-19}
 L^{\prime}(u_0) & = & {f_{y0}^{\prime}\ov\sqrt{f_{y0}}}  \ (\sin\th_1 K_1+\sin\th_2 K_2)\ ,
\nonumber\\
E'(u_0)& = &  {1\ov 2\pi} {f_{y0}^{\prime}} \sin\th_1 T_{p,q} \ (\sin\th_1 K_1+\sin\th_2 K_2)\ ,
\ea
which are indeed proportional to each other and where we have defined\footnote{In cases,
such as in Rindler space,
where the upper
limit of integration is finite, i.e. $u_{\rm max}=\L$, then there appears an additional term in
the expressions below given by $-\sqrt{F_i}/f'_y$ computed at $u=\L$ and also we replace infinity by $\L$
in the upper limit of integration.}
\be
 K_i=\int_{u_0}^{\infty}du\ \partial_u
\left({\sqrt{gf_y}\ov f^{\prime}_y}\right){1\ov(f_y-f_{yi})^{1/2}}\ ,\qq i=1,2\ .
\ee
The value of $u_{0c}$ for which \eqn{3-18} is satisfied is different
from $u_{0}=u_{0m}$
for which $L'(u_0)=E'(u_0)=0$. In fact, in the examples we will explicitly work out, it occurs
at smaller values, which implies that part of the upper branch of the solution is still perturbatively
stable although energetically not favorable.
However, the disconnected configuration becomes energetically favorable for
all positive values of the energy. Thus part of the
upper branch which is perturbatively stable is in fact metastable.
The fact that $u_{0c}\neq u_{0m}$ can be demonstrated quite
generally in the small coupling constant
limit for the most significant
case, namely for the quark-monopole system.
We may show that
\be
L'(u_0)= {f_{y0}^{\prime}\ov\sqrt{f_{y0}}}  \int_{u_0}^{\infty}du\ \partial_u
\left({\sqrt{gf_y}\ov f^{\prime}_y}\right){1\ov(f_y-f_{y0})^{1/2}}\ + {\cal O}(g_s) \ ,
\label{3-22}
\ee
with an analogous expression for $E'(u_0)$. From this we compute the leading term $u_{0m}^{(0)}$
in an expansion of the values $u_{0m}$ for which the maximum energy and length occur analogous to
\eqn{exppa}.
Using \eqn{th12},
\eqn{u12} and the general identity
\be
J_i(u_0)=2 K_i(u_0) + 2 {\sqrt{g_0}\ov f'_{y0} \cos\th_i }\ ,
\ee
proved by partial integration,
we may show that the zero-mode condition \eqn{3-18} becomes
\be
2 {\sqrt{g_0}\ov f'_{y0}} - \int_{u_0}^\infty du {\sqrt{g}\ov f_y}= {\cal O}(g_s)\ ,
\label{limgs}
\ee
from which one computes the leading order value of $u_{0c}$ in the expansion \eqn{exppa},
namely $u_{0c}^{(0)}$. Coming from different conditions we have that generically
$u_{0m}^{(0)}\neq u_{0c}^{(0)}$ and this
explicitly demonstrates that the value for $u_{0m}(g_s)$ found
by requiring maximum length and energy
is different than the value of $u_{0c}(g_s)$ found by requiring
the existence of a zero mode for the longitudinal
fluctuations which signals perturbative instability of the system.

\no
We note that the condition \eqn{limgs} can be directly obtained
from the boundary conditions \eqn{3-11} specialized to the zero mode solution for the longitudinal
fluctuations. In the derivation one should be careful and absorb a factor of $1/g_s$ into the overall
amplitude of the string 2.
In this way it becomes apparent that although in the small coupling limit
the string 2 becomes stiff and approximately straight, its
ultra small in strength fluctuations couple to those of the
string 1. Similar considerations can be given for the higher modes as well.
The above comments explain also why although in the limit $g_s\to 0$ (or $g_s\to \infty$)
the turning point $u_1\to u_0$ (or $u_2\to u_0$) and therefore the corresponding equation for the
fluctuations \eqn{cunv} becomes the same as the equation that appears in the
case of the single string for the quark-antiquark system \cite{ASS1}, the conditions
for the existence of the zero mode are different in the two cases. They are so due to the
different boundary conditions.

\no
Finally, we comment on whether or not the heavy dyon-dyon potentials as computed within supergravity
using the AdS/CFT correspondence violate some general principles regarding the strength and sign of
the corresponding force. For instance, for the quark-antiquark potential it turns out that the force should
always be attractive with strength an ever non-increasing function of the separation distance
\cite{concavity}. This is a concavity condition for the quark-antiquark potential and
implies that the upper branch in the $E\!-\! L$ plot should be disregarded, in
full agreement with the results of \cite{ASS1,ASS2} employing perturbative stability.
In our case, using \eqn{3-19}, we can show that
\ba
{dE\ov d L}& = & {T_{p,q} \sin\th_1\ov 2\pi} \sqrt{f_{y0}}\ ,
\nonumber\\
{d^2E\ov d L^2} &= & {T_{p,q} \sin\th_1\ov 4\pi} {f'_{y0}\ov \sqrt{f_{y0}} L'(u_0)}\ ,
\label{3-25}
\ea
which can be written in a symmetric way under an exchange of the strings 1 and 2 using the
$y$-component of the non-force condition.
The expressions \eqn{3-25} for the dyon-dyon potential are analogous to those
found in \cite{bs}
for the quark-antiquark potential.
Hence, the force is always attractive, but for the upper branch in the $E\!-\! L$ plot
in which $L'(u_0)>0$, has an increasing strength.
On the other hand we have found that part of the upper branch is perturbatively stable,
so that requiring
concavity seems to be a stricter condition than perturbative stability.
However, concavity is not a condition
that gauge theory analysis of Wilson loops for heavy quark-monopole pairs requires.
One can derive using
reflection-positivity, that the
energy of this pair is larger or equal than
the average of the energies of the quark-antiquark and
monopole-antimonopole systems.\footnote{We thank C. Bachas for providing this information.}
This condition is rather trivially satisfied in our case, provided we
use the physical lower branches of the two pairs.

\subsubsection{Angular zero modes}

For the angular zero modes we cannot explicitly write down the solution for the zero mode
due to the presence of the restoring force term
in the corresponding Sturm--Liouville equation in the third lines of
\eqn{3-6} and \eqn{jklo}.
Therefore, in this case we have to work out explicitly the result in the various
examples we will present. As it was shown in previous work \cite{ASS1,ASS2} one
can use approximate analytic methods in which the Schr\"odinger description of the differential equation
\eqn{3-7} plays an important role.

\section{ Examples of classical string junction solutions }
\label{sec-4}

In this section, we study first the behavior of the dyon-dyon
potentials emerging in Wilson-loop calculations for non-extremal
and multicenter D3-brane backgrounds.
Since we will work with the Nambu--Goto action, we need mention in
the expressions below only the metric and not the self-dual
five-form which is the only other non-trivial field present in our
backgrounds. At the end we work out the example of the Rindler space which captures
the behavior of black holes in various dimensions near the horizon.

\subsection{Non-extremal D3-branes}

We start by considering a background describing a stack of $N$
non-extremal D3-branes \cite{HoroStro}. In the field-theory limit the metric reads
\be
\label{4-1} ds^2 = {u^2 \ov R^2} \left[ - \left( 1 - {\m^4 \ov
u^4} \right)
 dt^2 + d \vec{x}_3^2 \right]
+ R^2 \left( {u^2 \ov u^4 - \m^4}\ du^2 + d \Omega_5^2 \right)\ ,
\ee
where the horizon is located at $u=\m$ and the corresponding
Hawking temperature
is $ T={\m \ov \pi R^2}$. This metric is just the direct product of
${\rm AdS}_5$--Schwarzschild with ${\rm S}^5$ and it is dual to
 $\cN=4$ SYM at finite temperature. For the calculations that
follow, it is convenient to switch to dimensionless variables by
rescaling all quantities using the parameter $\m$. Setting $u \to
\m u$ and $u_0 \to \m u_0$ and introducing dimensionless length
and energy parameters by
\be
\label{4-2}
L \to {1 \ov \m} L\ ,\qq E \to {\m \ov 2\pi} E\ ,
\ee
we see that all dependence on $\m$ drops out so that
 we may set $\m \to 1$ in what follows. The
functions in \eqn{2-2} depend only on $u$
(reflecting the fact that all values of $\th$ are equivalent) and
are given by
\ba
\label{4-3}
&&g(u) = 1\ ,\qq f_y(u) = (u^4-1)/R^4\ ,\qq f_x(u) = (u^4-1)/R^4\ ,\nonumber\\
&&f_\th(u) = {u^4-1 \ov u^2}\ ,
\qq h(u) = {u^4 \ov u^4 -1}\ .
\ea
In this case $u_{\rm min}=u_{\rm root}=1$, coinciding with the location of the
horizon.
For the dyon-dyon potential we find
that the separation length is given in terms of the junction point $u_0$ as
\ba
\label{4-5}
L &=& R^2 \left( \sqrt{u_1^4-1} \int_{u_0}^\infty {du \ov
 \sqrt{(u^4-1)(u^4-u_1^4)}}+(1\rightarrow 2)\right)=\ L_1+L_2\ ,\nonumber\\
L_i&=& R^2 {\sqrt{u_i^4-1}\ov 3u_0^3}
F_1\left({3\ov 4},{1\ov 2},{1\ov 2},{7\ov 4},{1\ov u_0^4},{u_i^4\ov u_0^4}\right)\ , \qq i=1,2\ ,
\ea
whereas the energy is given by
\ba
E&=&T_{p,q}{\cal E}_1+T_{p^{\prime},q^{\prime}}{\cal E}_2+T_{m,n}(u_0-1)\ ,
\nonumber\\
{\cal E}_i&=&\int_{u_0}^{\infty}du\left(\sqrt{{u^4-1\ov u^4-u_i^4}}-1\right)-(u_0-1)
\\
&=&
-u_0F_1\left(-{1\ov 4},-{1\ov 2},{1\ov 2},{3\ov 4},{1\ov u_0^4},{u_i^4\ov u_0^4}\right)+1\ ,\qq i=1,2\ .
\nonumber
\ea
In the above $F_1(a,b_1,b_2,c,z_1,z_2)$ is the Appell hypergeometric function
and $u_0 \geqslant u_i\geqslant 1$.
\no
From \eqn{2-14a} one finds that the turning points of the two strings are given by
\be
u_i=(u_0^4-(u_0^4-1)\cos^2\th_i)^{1/4}\ ,\qq i=1,2\ ,
\ee
where the angles $\th_i$ are given by \eqn{2-14b}, in terms of the strings' charges.
From \eqn{3-19} we have in this case
\ba
\label{4-6a}
&&L^\prime(u_0)= R^2 {4u_0^3\ov\sqrt{u_0^4-1}}\left(\sin\th_1K_1+\sin\th_2K_2\right)\ ,
\nonumber \\
&&K_i={3\ov28u_0^7}F_1\left({7\ov 4},{1\ov 2},{1\ov 2},{11\ov 4},{1\ov u_0^4},{u_i^4\ov u_0^4}\right)
-{1\ov12u_0^3}F_1\left({3\ov 4},{1\ov 2},{1\ov 2},{7\ov 4},{1\ov u_0^4},{u_i^4\ov u_0^4}\right)\ .
\ea
The function $L(u_0)$ has a single global maximum for any value of the string coupling.
Its location $u_{0m}$ depends on the string coupling and we obtain the following expansion
for the quark-monopole system
\be
u_{0m}\simeq 1.177 - 0.037 g_s + {\cal O}(g_s^2)\ .
\label{hsg}
\ee
For $L > L_{\rm max}$ , only the disconnected solution exists. For $L
< L_{\rm max}$, Eq. \eqn{4-5} has two solutions for $u_0$,
corresponding to a short and a long string and,
accordingly, $E$ is a double-valued function of $L$. The behavior described above is shown in
the plots of Fig. \ref{fig1} and we see that it is very similar to the behavior of the $L(u_0)$
and $E=E(L)$ for the heavy quark-antiquark system \cite{wilsonloopTemp}.
However, as we have pointed out and we will see
later in detail the upper branch, although energetically less favorable than the lower one, it
is not unstable in its entirety.
We note that the classical string junction in this case has also been examined in \cite{Park} where
the behavior of fig. \ref{fig1} was also noted.

\subsection{Multicenter D3-branes on a sphere}

We now proceed to the case of multicenter D3-brane distributions.
These distributions \cite{trivedi,sfet1} constricted as extremal limit of rotating
D3-brane solutions \cite{cy,rs} have been used in several studies for the Coulomb branch of
$\cN =4$ SYM
within the AdS/CFT correspondence, starting with the works of
\cite{bs,warn}. Here, we will concentrate on the particularly
interesting cases of uniform distributions of D3-branes on a three-sphere.

\no
The field-theory limit of the metric for $N$ D3-branes uniformly
distributed over a 3-sphere with radii $r_0$ reads
\ba
\label{4-22}
ds^2 &=& H^{-1/2} (- dt^2 + d \vec{x}_3^2 ) + H^{1/2}
{u^2-r_0^2 \cos^2\th \ov u^2-r_0^2}\ du^2
\nonumber\\
&+& H^{1/2}\left[(u^2-r_0^2\cos^2\th )d\th^2 + u^2 \cos^2\th d\Om_3^2
+ (u^2-r_0^2)\sin^2\th d \phi_1^2 \right]\ ,
\ea
where
\be
\label{4-23}
H = {R^4 \ov u^2 (u^2 - r_0^2 \cos^2 \th )}\ .
\ee
It is convenient to switch to dimensionless variables by
rescaling all quantities using the parameter $r_0$. Setting $u \to
r_0 u$ and $u_0 \to r_0 u_0$ and introducing dimensionless length
and energy parameters by
\be
\label{4-2new}
L \to {1 \ov r_0} L\ ,\qq E \to {r_0 \ov 2\pi} E\ ,
\ee
we see that all dependence on $r_0$ drops out so that
 we may set $r_0 \to 1$ in what follows., we write the
functions in \eqn{2-2}  as
\ba
\label{4-24}
&&g(u,\th) = {u^2 - \cos^2\th \ov u^2 - 1}\ ,
\qq f_y(u,\th) = u^2(u^2-\cos^2\th)/R^4\ ,
\nonumber\\
&& f_x(u,\th) = u^2(u^2-\cos^2\th)/R^4\ ,
\\
&&f_\th(u,\th)= u^2-\cos^2\th\ ,\qq h(u,\th) = {u^2 - \cos^2\th \ov u^2 - 1}\ .
\nonumber
\ea
Therefore the conditions \eqn{2-9} are satisfied
only for $\th_0=0$ and $\th_0=\pi/2$. We examine these two cases
in turn.

\subsubsection{The trajectory corresponding to $\th_0=0$}

In this case $u_{\rm min}=u_{\rm root}=1$ and
the integrals for the
dimensionless length and energy read
\ba
\label{4-26}
L &=& R^2 \left( u_1 \sqrt{u_1^2 - 1}
\int_{u_0}^\infty {du \ov u \sqrt{(u^2-1)(u^2-u_1^2)(u^2+u_1^2-1)}} +(1\to 2)\right)\nonumber\\
&=& R^2 \left({ u_1 k_1^{\prime} \ov u_1^2-1}
\left[ \elPi (\nu_1,k_1^{\prime 2},k_1) - \elF(\nu_1,k_1) \right]+(1\to 2)\right)
\ea
and
\ba
\label{4-27}
E &=& T_{p,q}{\cal E}_1+T_{p^{\prime},q^{\prime}}{\cal E}_2+T_{m,n}(u_0-1)\ ,
\nonumber\\
{\cal E}_i&=&\int_{u_0}^\infty du\left[ u \sqrt{u^2-1 \ov (u^2-u_1^2)(u^2+u_1^2-1)} - 1 \right] - (u_0-1)\ ,
\nonumber\\
&=&\sqrt{2u_i^2-1}[k_i^{\prime 2}({\bf K}(k_i)-{\bf F}(\mu_i,k_i))-({\bf E}(k_i)-{\bf E}(\mu_i,k_i))]
\\
&& +\sqrt{{(u_0^2-u_i^2)(u_0^2+u_i^2-1)\ov u_0^2-1}}+1\ ,
\qq i=1,2\ ,
\nonumber
\ea
\no
where ${\bf F}(\nu,k),{\bf E}(\nu,k)$ and $\elPi(\nu,\alpha,k)$
are the incomplete elliptic integrals of the
first, second and third kind respectively, while
${\bf K}(k),{\bf E}(k)$ and $\elPi(\alpha,k)$ are
the corresponding complete ones and
\ba
\label{4-28}
&&k_i={u_i \ov \sqrt{2u_i^2-1}}\ ,\quad k_i^{\prime}=\sqrt{1-k_i^2}\ ,\nonumber \\
&&\mu_i=\sin^{-1}\sqrt{{u_0^2-u_i^2\ov u_0^2-1}} ,\quad
\nu_i=\sin^{-1}\sqrt{{2u_i^2-1\ov u_0^2+u_i^2-1}}\ .
\ea

\no
From \eqn{2-14a} one finds that
\be
u_i=\displaystyle\sqrt{{1+\sqrt{\D_i}\ov 2}}\ , \qq \D_i=1+4u_0^2(u_0^2-1)\sin^2\th_i\ ,
\qq i=1,2 \ ,
\ee
where the angles $\th_i$ are given by \eqn{2-14b}.
The function $L(u_0)$ has a single global maximum
depending on the string coupling $g_s$. The situation is similar to the quark-antiquark
potential in this background \cite{bs} and
to the case of string junctions on black $D3$-branes in the present paper. The general behavior is shown in the
plots of Fig. \ref{fig1}.

\subsubsection{The trajectory corresponding to $\th_0=\pi/2$}

In this case $u_{\rm min}=1$, but $u_{\rm root}= 0$ and
the integrals for the
dimensionless length and energy read
\ba
\label{4-30}
L
&=& R^2  \left(u_1^2 \int_{u_0}^\infty {du \ov u \sqrt{(u^2-1)
(u^4-u_1^4)}}\ +(1\to 2)\right) \nonumber \\
&=& {R^2\ov \sqrt{2}}\left({\elPi(\nu_1,{1\ov 2},k_1)
-{\bf F}(\nu_1,k_1)\ov u_1}+(1\to 2)\right) ,
\ea
and
\ba
\label{4-31}
E &=& T_{p,q}{\cal E}_1+T_{p^{\prime},q^{\prime}}{\cal E}_2+T_{m,n}\sqrt{u_0^2-1} ,\nonumber \\
{\cal E}_i&=&\int_{u_0}^{\infty}{du u\ov\sqrt{u^2-1}}
\left({u^2\ov\sqrt{u^4-u_i^4}}-1\right)-\int_1^{u_0}{du u\ov\sqrt{u^2-1}}\ ,
\\
&=&\sqrt{2}u_i({\bf E}(\mu_i,k_i)-{\bf E}(k_i))+{u_i\ov\sqrt{2}}{\bf F}(\nu_i,k_i)
-\sqrt{(u_0^2-u_i^2)(u_0^2+u_i^2)\ov u_0^2-1}\ ,\nonumber
\ea
where now
\ba
\label{4-32}
&&k_i=\sqrt{{u_i^2+1 \ov 2 u_i^2}}\ ,\quad k^{\prime}=\sqrt{1-k^2}\ ,\nonumber\\
&&\mu_i=\sin^{-1}\left(\sqrt{{u_0^2-u_i^2\ov u_0^2-1}}\right)\ ,\quad
\nu_i=\sin^{-1}\left(\sqrt{{2u_i^2\ov u_0^2+u_i^2}}\right)\ .
\ea
\begin{figure}[!t]
\begin{center}
\begin{tabular}{cc}
\includegraphics[height=5.2cm]{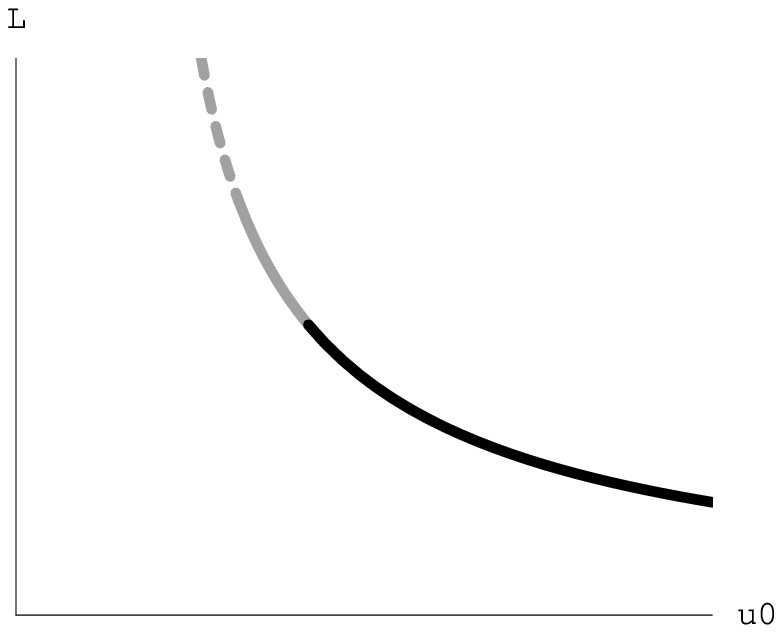}
&\includegraphics[height=5.2cm]{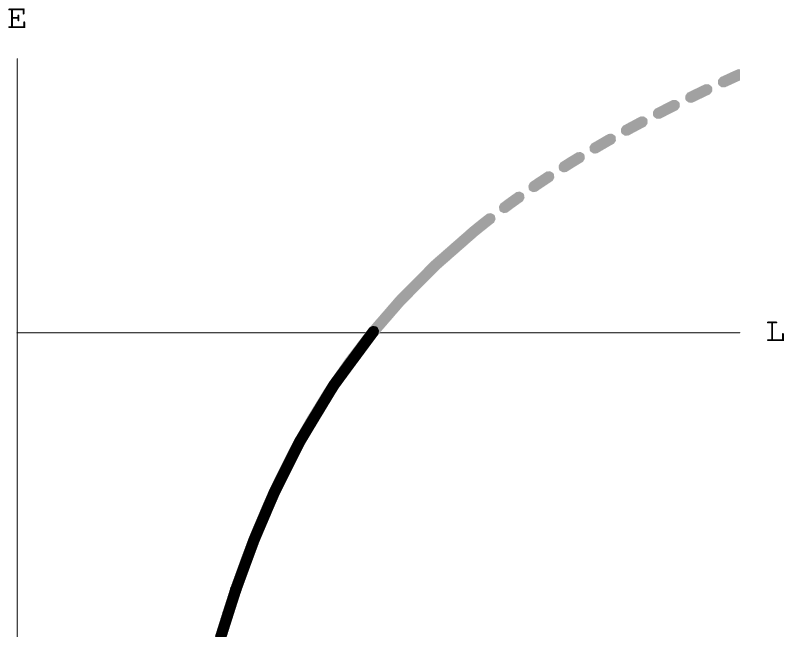}\\
(a) & (b)
\end{tabular}
\end{center}
\vskip -.5 cm \caption{Plots of $L(u_0)$ and $E(L)$ for cases
where there might be angular instabilities.
Note the appearance of a confining potential for cases where there are angular instabilities.
This covers, among our examples, the case of multicenter $D3$-branes on
a sphere for the $\th_0=\pi/2$ trajectory.
The meaning of the colors is as in fig. \ref{fig1}.}
\label{fig2}
\end{figure}
\no
From \eqn{2-14a} one finds that
\be
\label{4-33}
u_i=\displaystyle u_0\sqrt{\sin\th_i}\ , \qq i=1,2 \ ,
\ee
where   $\th_i$ angles are given by \eqn{2-14b}.
In this case $L(u_0)$ is a monotonously decreasing function which
approaches infinity as $u_0 \to 1$ and zero as $u_0 \to \infty$
and hence no maximal length exists, Eq. \eqn{4-30} has a single
solution for $u_0$ given the length $L$ and consequently
$E$ is a single-valued function of $L$.
This behavior is shown in Fig. \ref{fig2}.
This confining behavior is similar to what was found for the quark-antiquark system
in \cite{bs}.

\subsection{Rindler Space}

Our next example is based on the Rindler space, which
it is a portion of flat Minkowski space. In this space an observer
experiences the Unruh effect, perceiving the inertial vacuum as a
state populated by a thermal distribution of particles at a
temperature $T=\k/2\pi$, where $\k$ is the surface gravity.
Black hole solutions near the horizon behave like the Rindler space, so that the study
of string in this background serves more general purposes.
In \cite{Berenstein, ASS2}
the string configuration corresponding to an open string
with its two endpoints located at the same radius $u=\L$ (instead of $u\to \infty$)
we studied.
Moreover, we have proven that this problem is {\em exactly} equivalent
to the classical-mechanical problem of the
shape of a soap film suspended between two circular rings and we have also performed a
small fluctuations analysis around this classical configuration.

\no
The metric for Rindler space has the form
\be
\label{C-1} ds^2 = -\k^2 u^2 d t^2 + dy^2 + du^2 + \ldots\ ,
\ee
where $u$ is the radial direction with the Rindler horizon
corresponding to $u=0$ and $y$ is a generic spatial direction.
It is natural to extend the computation of \cite{ASS2} to the case of
sting junctions with the end points of strings 1 and 2 at $u=\L$ and the third
straight string stretching to $u=0$.
For convenience we pass to dimensionless units
\be
\label{C-2} u \to \L u\ ,\qq u_0 \to \L u_0\ ,\qq L \to \L L\ ,\qq
E \to {\k \L^2\ov 4\pi^2} E\
\ee
and then we use the formalism of section 2 which readily applies to this setup.
The length and the energy of the string are determined by Eqs.
\eqn{2-12} and \eqn{2-13} (without the subtraction term for the energy and the
upper limit in the integration changed to 1) which for
our case read
\be
\label{C-3} L = \int_{u_0}^1 du{u_1\ov{\sqrt{u^2-u_1^2}}}\ + (1\to 2)
= u_1\left(\cosh^{-1}{1\ov u_1}-\cosh^{-1}{u_0\ov u_1}\right)+(1\to 2)
 ,
\ee
and
\ba
\label{C-4} E &=&T_{p,q}{\cal E}_1+T_{p^{\prime},q^{\prime}}{\cal E}_2+T_{m,n}\pi u_0^2\ ,
  \\
 {\cal E}_i &=&\int_{u_0}^1 {d u \, u^2 \ov\sqrt{u^2-u_i^2}}=
\pi \left( \sqrt{1-u_i^2} + u_i^2 \cosh^{-1}
{1 \ov u_i}-u_0\sqrt{u_0^2-u_i^2}-u_i^2\cosh^{-1}{u_0 \ov u_i}\right)\ . \nonumber
\ea
From \eqn{2-14a} one finds that
\be
u_i=\displaystyle u_0\sin\th_i\ , \qq i=1,2 \ ,
\ee
 where the  angles $\th_i$ are given by \eqn{2-14b}.
The function $L(u_0)$ has a single global maximum. Its location,
depends on the string coupling $g_s$. For small values of the latter we get the expansion
\be
u_{0m}\simeq 0.552- 0.450 g_s + 0.479 g_s^2 -0.100 g_s^3 + {\cal O}(g_s^4)\ .
\label{4-25}
\ee
The situation is similar to that depicted in the plots of fig. \ref{fig1}.

\section{Examples of stability analysis}

In this section we apply the stability analysis developed in
section 3 to the string solutions reviewed in section 4. We find
that for the non-extremal D3-brane, the sphere with $\th_0=0$ and the Rindler space
we have a longitudinal instability corresponding to the upper
branch of the energy curve,
for the sphere with $\th_0=\pi/2$ we have angular instabilities
towards the IR, even though the potential has a single branch.

\subsection{The conformal case}

For completeness we consider first the conformal case, corresponding to the $\mu \to 0$ or
$r_0 \to 0$ limit of any of the above solutions. The classical
string junction solution
has been constructed in \cite{Minahan}.
Since there is no $\th$-dependence, the $\d\th$ fluctuations are
 equivalent to the transverse $\d x$-fluctuations and therefore stable.
On the other hand for the longitudinal fluctuations we have to solve \eqn{3-18}.
From \eqn{2-14a} one finds that
\be
u_i=\displaystyle u_0\sqrt{\sin\th_i}\ ,\qq i=1,2 \ ,
\ee
where the angles $\th_i$ are given by \eqn{2-14b}. Moreover
we compute
\be
 J_i(u_0)={1\ov 3u_0^3}\ {_2F_1}\left({3\ov 4},{3\ov 2},{7\ov 4},{u_i^4\ov u_0^4}\right)\ .
\ee
Using the above, we may check that \eqn{3-18} has no solution for any value of string coupling $g_s$,
thus we verify that indeed string junctions are stable in the conformal case.

\subsection{Non-extremal D3-branes}

We proceed next to the case of non-extremal D3-branes, where we
recall that the potential energy is a double-valued function of
the separation length.
As before, since there is no $\th$-dependence, the $\d\th$ fluctuations are equivalent to the
$\d x$-fluctuations and stable. For the longitudinal fluctuations we have to solve \eqn{3-18}, where
\ba
 J_i(u_0)={1\ov 3u_0^3}\
F_1\left({3\ov 4},-{1\ov 2},{3\ov 2},{7\ov 4},{1\ov u_0^4},{u_i^4\ov u_0^4}\right)\ ,
\qq i=1,2\ .
\ea
Equation \eqn{3-18} has a solution for any value of string coupling $g_s$. This
value $u_{0\rm c}$,
is smaller than the value $u_{0m}$ in which the maximum
of the length occurs.
In the limit of small coupling constant for the quark-monopole
system we have shown that we can approximate \eqn{3-18} by the condition \eqn{limgs} which in
our case becomes
\be
{}_2F_1\left({3\ov 4},1,{7\ov 4},{1\ov u_0^4}\right)-{3\ov 2}={\cal O}(g_s)\ .
\ee
Its solution gives the
leading order result for the critical value in which instability occurs
\be
u_{0c}\simeq  1.117 -0.166 g_s + {\cal O}(g_s^2)\ ,
\ee
where we have included the first correction as well.
This expansion is different than
the expansion of $u_{0m}$ in \eqn{hsg}.
Thus the whole upper branch is not entirely unstable, a
behavior not expected from energetic considerations.
Let's compare the maximum length of the screened interaction for the cases of the
 quark-antiquark
and the quark-monopole interactions.
As $g_s$ varies from small to large values the maximum length is
$L^{\rm qm}_{\rm max}\simeq (0.425\pm 0.005) R^2$, whereas for the quark-antiquark pair
$L^{\rm q\bar q}_{\rm max}\simeq 0.869 R^2\simeq 2 L^{\rm qm}_{\rm max}$. We see that the quark-monopole
pair is screened twice as strong as the quark-antiquark pair. This is a general feature in all of the
examples we encounter in this paper.

\subsection{Multicenter D3-branes on a sphere}

The results of the stability analysis for the two orientations for the sphere
distribution are presented below.

\subsubsection{The trajectory corresponding to $\th_0=0$}

Angular fluctuations are stable due to the fact that, as in \cite{ASS1}, we may transform,
for each string separately,
the problem into a Schr\"odinger equation with a positive potential for all values of $u_i$.
For the longitudinal fluctuations we have to solve \eqn{3-18}, where
\ba
J_i(u_0)=\int_{u_0}^\infty du{u\sqrt{u^2-1}\ov (u^2(u^2-1)-u_i^2(u_i^2-1))^{3/2}}\ ,
\qq i=1,2\ .
\ea
This has a solution for all values of the string coupling.
In the limit of small string coupling for the quark-monopole
system the condition \eqn{limgs} reads explicitly
\be
{2u_0^2\ov 2u_0^2-1} -{u_0\ov 2}\ln\left(u_0+1\ov u_0-1\right) ={\cal O}(g_s)\ ,
\ee
giving the
leading order result for the critical value in which instability occurs
\be
u_{0c}\simeq  1.084 + {\cal O}(g_s)\ .
\ee
Thus, part of the upper branch of the solution is
perturbatively stable as in the case of the string junction based on the non-extremal D3-brane solution.
As before, we compare the maximum length of the screened interaction for the cases of the quark-antiquark
and the quark-monopole interactions.
As $g_s$ varies from small to large values the maximum length in the quark-monopole interaction is
$L^{\rm qm}_{\rm max}\simeq (0.425\pm 0.050) R^2$, whereas for the quark-antiquark pair
 $L^{\rm q\bar q}_{\rm max}\simeq 1.002 R^2\simeq 2.3 L^{\rm qm}_{\rm max}$.

\subsubsection{The trajectory corresponding to $\th_0=\pi/2$}

Longitudinal fluctuations are stable since an
explicit check of \eqn{3-18} shows that it has no solution for any value of string coupling $g_s$.
Alternatively, we may transform the problem  into a Schr\"odinger problem as in
\cite{ASS1} with
an everywhere positive potential.
For the angular fluctuations the potential is just constant
\be
V_\th =-1 \ , \qq i=1,2\ .
\ee
The change of variables for the two string is
\be
 z_i(u)={1\ov u_i\sqrt{2}}\ {\bf F}(\nu_i,k_i)\ ,\quad  i=1,2\ ,
\ee
where $\nu_i$ and $ k_i$ are given by
\ba
\nu_i=\sin^{-1}\left(\sqrt{{2u_0^2 \sin\th_i\ov u^2+u_0^2 \sin\th_i}}\right)\ ,
\qq k_i=\sqrt{{u_0^2\sin\th_i+1\ov 2u_0^2\sin\th_i}}\ ,\qq i=1,2\ .
\ea
The zero mode of the Schr\"odinger equation for the $i=1,2$ strings read
\ba
\d \th_i(z_i)=c_i\sin z_i\ ,\quad z_i\in[0,\zeta_{i}]\ ({\rm as}\ u\in (\infty,u_0])\ ,
\quad \zeta_{i}=z_i(u_0) \ ,
\ea
where $c_i$ are multiplicative constants. The straight string 3 plays a r\^ole in the boundary
conditions, the reason being that the mass term is present in the equation for the angular
fluctuations even for the zero mode.
For the straight string we have two linear independent solutions
$\displaystyle\sin z$ and $\cos z$, where the appropriate change of variable is
\be
z={\pi\ov 2}-\sin^{-1}{1\ov u} \ .
\ee
We examine in detail the case where only $\sin z$ is used
and comment on the general linear combination later.
From the matching conditions in \eqn{3-18ab} one finds the following condition
\ba
\sin\th_1\sin \zeta_1 \cos \zeta_2 +\sin\th_2\sin  \zeta_2 \cos \zeta_1
- {\sin(\th_1+\th_2)\ov u_0(u_0^2-1)}\sin \zeta_1\sin \zeta_2=0\ .
\label{5-14}
\ea
For the quark-monopole case we may use that $\th_1+\th_2=\pi/2$, to simplify it
a bit. Furthermore, in the small string coupling limit \eqn{5-14}, becomes
\be
\tan \left({1\ov 2 u_0} + {1\ov 2 \sqrt{u_0^2-1}}\right)-u_0(u_0^2-1) ={\cal O}(g_s)\ ,
\ee
which is a more explicit and easy to handle
 transcendental equation. Hence, the solution for the leading
order critical values for which instability of the angular perturbations occur
is
\be
u_{0 c}\simeq 1.378 + {\cal O}(g_s)\ .
\ee
The confining behavior turns out to be unstable since
\eqn{5-14} has a solution for any value of the string coupling $g_s$.
As $g_s$ varies from small to large values,
 the maximum length of the quark-monopole interaction is
$L^{\rm qm}_{\rm max}\simeq (0.560\pm 0.040) R^2$, whereas for the quark-antiquark pair
 $L^{\rm q\bar q}_{\rm max}\simeq 1.700 R^2\simeq 3 L^{\rm qm}_{\rm max}$.

\no
Let' s mention that, had we used the general solution
of the Scr\"odinger equation, that is $\cos (z +\varphi)$,
we would have obtained similar results for any
phase with the exception of $\varphi =0$ for which case no instability occurs.
Hence the above considerations and results are
quite generic.

\no
Finally let's note that, in order keep the discussion at a reasonable length, we refrained from
explicitly presenting the details for the other important uniform distribution of D3-branes, namely, that
on a disc. We mention the end result:
For $\th_0=\pi/2$ the solution is stable against all types of small fluctuations,
while for $\th_0=0$ it is stable against longitudinal fluctuations and unstable against angular
ones after a certain separation length.
This behavior reveals the expected screening behavior which moreover is practically independent
of the orientation of the string and similar to that for the case of the
quark-antiquark potential studied in full detail in \cite{ASS1}.

\subsection{Rindler Space}

Finally, we consider the case of the Rindler space.  Fluctuations of the single string solution corresponding to
the heavy quark-antiquark potential have been considered in \cite{ASS2,Berenstein}.
Turning to our case, we see from the form of the metric only the
longitudinal fluctuations are relevant for our discussion, so that
we have to solve \eqn{3-18}, where
\ba
 J_i(u_0)={1\ov \kappa}\left(\cosh^{-1}{1\ov u_i}-\cosh^{-1}{u_0\ov u_i}-{1\ov\sqrt{1-u_i^2}}
+{u_0\ov\sqrt{u_0^2-u_i^2}}\right)\ .
\ea
Equation \eqn{3-18} has a solution for all values of the string coupling $g_s$ and we have the
same behavior as in $\d y$
fluctuations in the non--extremal D3--branes.
For small values of the string coupling latter we get the expansion
\be
u_{0c}\simeq 0.368 
- 0.214 g_s + 0.218 g_s^2 -0.058 g_s^3+ {\cal O}(g_s^4)\ .
\ee
We see that it is different from the expansion of $u_{0m}$ in \eqn{4-25}
corresponding to the maximum of the length and the energy. Hence, the behavior is similar
to that represented in fig. \ref{fig2}.

\section{Discussion}

In this paper we have constructed string
junctions on general backgrounds and used them to compute the interaction energy of heavy
dyons at strong coupling within the AdS/CFT correspondence. The most important case is that of
the quark-monopole pair for which we paid particular attention.
Then we turned into the perturbative stability analysis of these solutions which is important to
distinguish the physically relevant regions in which the potentials can be trusted.
We formulated and further investigated general conditions for the existence of instabilities.
Starting with the conformal case we indeed verified that there are no instabilities at the
conformal point of $\cN=4$ SYM, as expected. Nevertheless, we emphasize here
that in the $\b$-deformed conformal solution of \cite{LM},
as well as for all backgrounds that asymptote this, there is an instability for
certain angular fluctuations.
The reason is identical to the case of a single string for the quark-antiquark potential
that was found in \cite{ASS2} (see section 4.3.1) and occurs for values of the real deformation parameter
$\s$ (in the standard notation) larger than a certain critical value (we remark
that the classical string junction of the $\b$-deformed background of the disc D3-brane distribution has been
considered in \cite{Ahn} and should suffer from this instability).

\no
We found that part of the branches of the potential energy versus length
that could be disregarded on energetic considerations can be perturbatively stable.
The reason is traced to the fact that the matching conditions at the junction point allow
for interaction of all strings in the junction that enhances its stability.
This is unlike the findings in \cite{ASS1,ASS2} for the quark-antiquark potential
according to which the perturbative instability renders energetically unfavorable branches as
unstable.
We pointed out that this result is not in conflict with general considerations on the quark-monopole potential.
For
instance, there is no concavity condition as is in the case of the quark-antiquark potential.
We found that the confining behavior of the dyon-potential in the Coulomb
branch of the $\cN=4$ SYM with vacuum expectation values distributed uniformly on a sphere, is unstable. This
is similar to the result of \cite{ASS1} for the seemingly confining behavior of the quark-antiquark
potential for the same distribution. It is quite remarkable that demanding perturbative
stability of the string configurations renders
as physically unacceptable the confining behavior, which indeed is
not expected from gauge theory considerations.

\no
Our general formalism was based on a diagonal class of metrics \eqn{2-1}.
It can be extended to
supergravity backgrounds in which off-diagonal metric elements appear and in which additional
background form fields couple in the Wess--Zumino part of the string actions in the junction,
in analogy with \cite{ASS2} for the quark-antiquark case.
In particular, this will cover, within the AdS/CFT correspondence,
interactions of moving dyons in hot quark-gluon plasmas.
We expect that also in this case the associated energy will have two branches with the upper one
being only partly unstable.
The behavior found in \cite{LRW}, namely that the maximum length behaves for high enough velocity as
$L_{\rm max}\simeq 0.743 (1-v^2)^{1/4} R^2$ is expected to persists with the overall numerical
coefficient replaced by, roughly, its half.

\no
It will be interested to extend our analysis to junctions involving more than three
strings where we believe similar phenomena with the triple-junctions considered here will be found.
A more interesting generalization is to consider systems of multiple external particles
at finite temperature and examine to what extend
 they prefer, under small perturbations to split, with raising temperature,
into smaller systems. This possibility was argued for, based on energy considerations, in
\cite{polyulf} who examined in detail the interaction energy for the system of a
quark, a monopole and a dyon.

\no
Finally, it has been proposed that
$F$- and $D$-superstrings could have been produced at the early Universe,
then expanded at a cosmic size and
that their network properties by forming junctions make them distinguishable \cite{CoMyPol}
from gauge theory cosmic strings (see, for instance, \cite{vilenkinBook}).
The no-force condition
for a stationary junction
to form, gives rise to a locally flat space time  \cite{Brandenberger},
much-like the case of single cosmic strings \cite{Vilenkin}. Small fluctuations of cosmic
string junctions on flat spacetime have been analyzed \cite{kibble}
in closed analogy with corresponding work
on superstring junctions (see, for instance, \cite{vilenkinBook}). In view of the above remarks
it will be very interested to study the formation and stability
of cosmic size string junctions in cosmological backgrounds, by adapting the results and
techniques of the present paper, to such cases.

\vskip 0.5cm

\centerline{ \bf Acknowledgments}

\no
We would like to thank S. Avramis for useful discussions.
We acknowledge support provided through the European
Community's program ``Constituents, Fundamental Forces and
Symmetries of the Universe'' with contract MRTN-CT-2004-005104,
the INTAS contract 03-51-6346 ``Strings, branes and higher-spin
gauge fields'', the Greek Ministry of Education programs $\rm \P
Y\Th A\G OPA\S$ with contract 89194. In addition, K.~Siampos acknowledges support
provided by the Greek State Scholarship Foundation (IKY).

\appendix

\section{An analog from classical mechanics}
Longitudinal string fluctuations exhibit a behavior which is not expected from the energetics
of the configurations.
However, this behavior can be found in the classical mechanics problem
of determining
the shape of a thin soap film stretched between two rings appropriately,
modified to serve our purposes.
We briefly review first the solution of the unmodified classical problem which is a catenary and
can be found in standard textbooks \cite{variations}.
In the notation of the present paper and
of the appendix of \cite{ASS1} the cylindrically symmetric solution reads
\be
r(z)=u_0 \cosh{z\ov u_0}\ ,\qq -{L\ov 2}\leqslant z \leqslant {L\ov 2}\ ,
\label{soopp}
\ee
in terms of the integration constant parameter $u_0$.
The potential energy and length of the soap film
 are given by the following expressions
\ba
&&L=2u_0\cosh^{-1}{1\ov u_0}\ ,
\nonumber\\
&&E=2\pi\left(\sqrt{1-u_0^2}+u_0^2\cosh^{-1}{1\ov u_0}\right)\ ,\quad 0\leqslant u_0
\leqslant 1\ .
\ea
Hence, $E(L)$ is a double-valued function of $L$ with the energetically
favorable branch having lower energy for a given distance, corresponding to a shallow
catenary, whereas the upper branch corresponds to a deep catenary.
There is also the Goldschmidt solution, with two
disconnected soaps in the two rings with energy $E=2\pi$.
Thus we have the same qualitative behavior as in the classical solution of the non--extremal D3--branes,
depicted in fig. 2.  The
value of $u_0$ at the turning point is $u_{0m}\simeq 0.552$ leading to the maximal length of $L_m\simeq1.325$.
The  Goldschmidt solution becomes energetically favorable at a higher value of $u_0\simeq 0.826$,
corresponding to a smaller value of $L\simeq 1.055$.
The perturbative small fluctuation analysis has been performed in \cite{durand} and in \cite{ASS1}.
The small fluctuations stability analysis of this solution gives the expected result
that the upper,
energetically
unfavorable,  branch
is perturbatively unstable.

\no
We would like to add an extra
term to the quadratic fluctuations (see Eq.(A-1) in appendix A of \cite{ASS1}) that, to leading order,
would not affect the energy of the solutions and then check whether part of the upper branch
becomes perturbatively stable due to this modification of the action.
In order to fulfill the above requirement imagine that we first let the soap film assume its shape
given by \eqn{soopp}
and then we turn on the perturbation which we take it to be quadratic in
the normal fluctuations of the soap film surface. The normal fluctuations decouple
from the two tangential perturbations
which moreover are trivial \cite{ASS1}. This will
not affect the classical solution and its energy, but will certainly modify the stability analysis.
There are several such terms that we may add. Here we choose to present
the two most natural ones.

\no
In the first we add a term to the potential that acts only at the minimum of the catenary at $z=0$
\be
V= {\k u_0\ov 2}(\d\eta)^2\d(z)  \ ,\qq k\geqslant 0\ ,
\label{A-3}
\ee
where the factor $u_0$ is introduced for convenience.
For the normal perturbations, we define
$\displaystyle u={z/u_0}$, we separate variables according to
\ba
\d\eta(t,u,\phi)=\Phi(u)e^{-i\Omega t}e^{i m\phi}\ ,\quad m=0,\pm1,\pm2,\dots\ ,
\ea
and eventually we end up with the Sturm-- Liouville equation
\ba
\label{A-4}
-{d^2\Phi\ov du^2}+\left(-{2\ov\cosh^2u}+m^2+\kappa \d(u) \right)\Phi=\omega^2\cosh^2u\Phi\ ,\qq
\omega=u_0\Omega \ ,
\ea
subject to the following boundary conditions
\ba
\label{A-5}
\Phi\left(\pm{L\ov2u_0}\right)=0 \ ,
\ea
representing fixed endpoints.
Equation \eqn{A-4} can not be easily solved for non-vanishing values of $\omega$, but we can
obtain useful information for our stability problem by considering the zero--mode problem having  $\omega=0$.
The transformation $x=\tanh u$ turns \eqn{A-4} to an associated Legendre equation with the
general solution given by a linear combination of $P_1^{m}(\tanh u)$ (they vanish identically unless $m=0,\pm 1$)
and $ Q_1^{m}(\tanh u)$,
with different coefficients to the regions left and right of $z=0$.
However, we know from the work in \cite{ASS1} that in the absence of the extra term (for $\k=0$),
only for $m=0$ a zero mode exists for the
value $u_0\simeq 0.552$ we mentioned above. Since the modification is by an repulsive $\d$-function term the
possibility of having zero mode solutions in the presence of it for $|m|\geqslant 1$,
is definitely excluded. Hence we concentrate
to the case with $m=0$.
The continuity of the solution at $z=0$,
the discontinuity of its derivative, as read off from \eqn{A-4}, and the boundary
conditions \eqn{A-5}, give a linear homogeneous system for the
four independent coefficients  which in order to have non-trivial solution
should have a vanishing determinant resulting to the condition
\ba
\label{A-7}
\k  P_1(a) =2 Q_1(a)\ ,
\quad \Longrightarrow\quad \k a + 2 = a \ln\left(1+a\ov 1-a\right)\ ,
\qq a=\sqrt{1-u_0^2}\ .
\ea
This equation has a  solution for all $\k\geqslant 0$, which varies monotonously from $u_{0c}\simeq 0.552$
to 0, as $\k\in [0,\infty)$. Importantly, the product $u_{0c} \k$ and therefore the term \eqn{A-3}
we have added, remain finite.
For small and large values of $\k$ we have the behaviors
\ba
u_{0c} \simeq 0.552 - 0.160 \kappa  + {\cal O}(\kappa^2)\
\ea
and
\ba
u_{0c} = 2 e^{-\kappa/2-1} + {\cal O}(e^{-\kappa})\ .
\ea

\no
An alternative to \eqn{A-3} potential term is
\be
V ={k u_0\ov 2}(\d\eta)^2 \ , \qq k\geqslant 0\ ,
\ee
where the factor $u_0$ is, as before, introduced for convenience.
In the Sturm--Liouville equation \eqn{A-4} we replace $\kappa \d(u)\to u_0 k$.
The same transformation $x=\tanh u$ gives an associated Legendre equation with the
general solution given by a linear combination of $P_1^{\widetilde{m}}(\tanh u)$
and $ Q_1^{\widetilde{m}}(\tanh u)$, where $\widetilde{m}=\sqrt{m^2+u_0k}$.
For identical reasons as before we concentrate to the case with $m=0$ and in fact refining the
argument we should have that $\widetilde{m}<1$  in order to find a normalizable
zero mode.
Imposing the boundary condition \eqn{A-5} and demanding that the determinant of the resulting
linear homogeneous system for two independent coefficients
has non-trivial solutions, we obtain the condition
\ba
\label{A-6}
P_1^{\sqrt{u_0k}}(a) Q_1^{\sqrt{u_0k}}(-a)=P_1^{\sqrt{u_0k}}(-a)Q_1^{\sqrt{u_0k}}(a)\ ,\qq a=\sqrt{1-u_0^2}\ .
\ea
This can only be solved numerically and as before the product $u_{0c} k$
remains finite.
For small and large values of $k$ we have the behaviors
\ba
u_{0c}\simeq 0.552 - 0.102 k^4  + {\cal O}(k^8)\
\ea
and
\ba
u_{0c} = {1\ov k} -{2\ov k^3}  + {\cal O}(1/k^5)\ .
\ea

\no
In both of the above examples we see that the critical value $u_{0c}$ can get arbitrarily
small, which implies that the entire upper, energetically less favorable branch corresponding
to the deep catenary, can become perturbatively stable.


\begin{thebibliography}{99}




\renewcommand{\baselinestretch}{1}
\normalsize

\bibitem{AhaSoYa}
  O.~Aharony, J.~Sonnenschein and S.~Yankielowicz,
  Nucl. Phys. {\bf B474} (1996) 309, {\tt hep-th/9603009}.

\bibitem{Schwarz}
  J.H.~Schwarz,
  Nucl. Phys. Proc. Suppl.\  {\bf 55B} (1997) 1, {\tt hep-th/9607201}.

\bibitem{Dasgupta}
  K.~Dasgupta and S.~Mukhi,
  Phys. Lett. {\bf B423} (1998) 261, {\tt hep-th/9711094}.

\bibitem{Sen}
  A.~Sen,
  JHEP {\bf 9803} (1998) 005, {\tt hep-th/9711130]}.

\bibitem{ReyYee}
  S.J.~Rey and J.T.~Yee,
  Nucl. Phys. {\bf B526} (1998) 229, {\tt hep-th/9711202}.

\bibitem{KroghLee}
  M.~Krogh and S.~Lee,
  Nucl. Phys. {\bf B516} (1998) 241, {\tt hep-th/9712050}.

\bibitem{Matsuo}
  Y.~Matsuo and K.~Okuyama,
  Phys. Lett. {\bf B426} (1998) 294, {\tt hep-th/9712070}.

\bibitem{Bergman}
  O.~Bergman,
  Nucl. Phys. {\bf B525} (1998) 104, {\tt hep-th/9712211}.

 \bibitem{Gabe}
   M.R.~Gaberdiel and B.~Zwiebach,
   Nucl. Phys. {\bf B518} (1998) 151, {\tt hep-th/9709013}.

\bibitem{CaTho}
  C.G.~Callan and L.~Thorlacius,
  Nucl. Phys. {\bf B534} (1998) 121, {\tt hep-th/9803097}.

\bibitem{adscft} J.M.~Maldacena,
Adv.\ Theor.\ Math.\ Phys.\ {\bf 2} (1998) 231, Int.\ J.\ Theor.\ Phys.\  {\bf 38} (1999) 1113,
{\tt hep-th/9711200}.\hfill\break
S.S.~Gubser, I.R.~Klebanov and A.M.~Polyakov,
Phys.\ Lett.\ {\bf B428} (1998) 105, {\tt hep-th/9802109}.\hfill\break
E.~Witten,
Adv.\ Theor.\ Math.\ Phys.\ {\bf 2} (1998) 253, {\tt hep-th/9802150}
and
Adv.\ Theor.\ Math.\ Phys.\  {\bf 2} (1998) 505,
{\tt hep-th/9803131}.

\bibitem{Minahan}
 J.A.~Minahan,
  Adv.\ Theor.\ Math.\ Phys.\  {\bf 2} (1998) 559, {hep-th/9803111}.

\bibitem{ASS1}
  S.D.~Avramis, K.~Sfetsos and K.~Siampos,
  Nucl. Phys. {\bf B769} (2007) 44,\hfill\break {\tt hep-th/0612139}.

\bibitem{ASS2}
  S.D.~Avramis, K.~Sfetsos and K.~Siampos,
Nucl. Phys. {\bf B793} (2008) 1, \hfill\break {\tt arXiv:0706.2655 [hep-th]}.

\bibitem{maldaloop} J.M.~Maldacena,
Phys. Rev. Lett. {\bf 80} (1998) 4859, {\tt
hep-th/9803002}.\hfill\break
S.J.~Rey and J.T.~Yee,
Eur. Phys. J. {\bf C22} (2001) 379, {\tt hep-th/9803001}.

\bibitem{bs}
A.~Brandhuber and K.~Sfetsos,
Adv. Theor. Math. Phys. {\bf 3} (1999) 851,\hfill\break {\tt
hep-th/9906201}.

\bibitem{concavity}
B.~Baumgartner, H.~Grosse and A.~Martin,
Nucl.\ Phys.\ {\bf B254} (1985) 528.\hfill\break
C.~Bachas,
Phys.\ Rev.\ {\bf D33} (1986) 2723.

\bibitem{HoroStro}
  G.T.~Horowitz and A.~Strominger,
  Nucl. Phys. {\bf B360} (1991) 197.

\bibitem{wilsonloopTemp}
S.J.~Rey, S.~Theisen and J.T.~Yee,
Nucl.\ Phys.\ {\bf B527} (1998) 171, {\tt hep-th/9803135}.
\hfill\break
A.~Brandhuber, N.~Itzhaki, J.~Sonnenschein and S.~Yankielowicz,
Phys.\ Lett.\ {\bf B434} (1998) 36, {\tt hep-th/9803137}
and
JHEP {\bf 9806} (1998) 001, {\tt hep-th/9803263}.

\bibitem{Park}
  D.K.~Park,
  Nucl. Phys. {\bf B618} (2001) 157, {\tt hep-th/0105039}.

\bibitem{trivedi} P.~Kraus, F.~Larsen and S.P.~Trivedi,
JHEP {\bf 9903} (1999) 003, {\tt hep-th/9811120}.

\bibitem{sfet1} K.~Sfetsos,
JHEP {\bf 9901} (1999) 015, {\tt hep-th/9811167}.

\bibitem{cy}
M.~Cvetic and D.~Youm, Nucl. Phys. {\bf B477} (1996) 449, {\tt
hep-th/9605051}.

\bibitem{rs}
J.G.~Russo and K.~Sfetsos,
Adv. Theor. Math. Phys.  {\bf 3} (1999) 131, {\tt hep-th/9901056}.

\bibitem{warn} D.Z.~Freedman, S.S.~Gubser, K.~Pilch and N.P.~Warner,
JHEP {\bf 0007} (2000) 038, {\tt hep-th/9906194}.


\bibitem{Berenstein}
  D.~Berenstein and H.J.~Chung,
  {\it Aspects of open strings in Rindler Space}, \hfill\break
{\tt arXiv:0705.3110 [hep-th]}.

 \bibitem{LM}
 O.~Lunin and J.M.~Maldacena,
 JHEP {\bf 0505} (2005) 033, {\tt hep-th/0502086}.

\bibitem{Ahn}
  C.~Ahn,
  Phys. Lett. {\bf B641} (2006) 481, {\tt hep-th/0605012}.


\bibitem{LRW}
H.~Liu, K.~Rajagopal and U.A.~Wiedemann,
Phys. Rev. Lett. {\bf 98} (2007) 182301, {\tt hep-ph/0607062},
{\tt hep-ph/0607062}.

\bibitem{polyulf}
  U.H.~Danielsson and A.P.~Polychronakos,
  Phys. Lett. {\bf B434} (1998) 294, {\tt hep-th/9804141}.


\bibitem{CoMyPol}
  J.~Polchinski,
  AIP Conf. Proc.  {\bf 743} (2005) 331
  [Int. J. Mod. Phys. {\bf A20} (2005) 3413]
  [arXiv:hep-th/0410082].
\hfill\break
  E.J.~Copeland, R.C.~Myers and J.~Polchinski,
  JHEP {\bf 0406} (2004) 013, {\tt hep-th/0312067}.



\bibitem{vilenkinBook}A. Vilenkin and E. Shellard, {\it Cosmic strings and other
gravitational defects}, Cambridge University Press, 1994.\hfill\break
M.B.~Hindmarsh and T.W.~B.~Kibble,
  Rept. Prog. Phys. {\bf 58} (1995) 477, {\tt hep-ph/9411342}.

\bibitem{Brandenberger}
  R.~Brandenberger, H.~Firouzjahi and J.~Karouby,
  {\it Lensing and CMB Anisotropies by Cosmic Strings at a Junction}, {\tt arXiv:0710.1636 [hep-th]}.


\bibitem{Vilenkin}
  A.~Vilenkin,
  Phys. Rev. {\bf D23} (1981) 852.


\bibitem{kibble}
  E.J.~Copeland, T.W.~B.~Kibble and D.A.~Steer,
  Phys. Rev. Lett.  {\bf 97} (2006) 021602, {\tt hep-th/0601153}
and
  Phys. Rev. {\bf D75} (2007) 065024, {\tt hep-th/0611243}.


 \bibitem{variations}
 C.~Caratheodory, {\em Calculus of Variations and Partial
 Differential Equations of the First Order}, American Mathematical
 Society, 1999.\hfill\break G.A.~Bliss, {\em Calculus of
 Variations}, Chicago, IL: Open Court, 1925. \hfill\break
 R.~Courant and H.~Robbins, {\em What is Mathematics?}, Oxford
 University Press, Oxford, 1941. \hfill\break G.B.~Arfken and
 H.J.~Weber, {\it Mathematical Methods for Physicists}, 6th
 edition, 2005, Elsevier Academic Press.

\bibitem{durand}
L.~Durand,
Am. J. Phys. {\bf 49} (1981) 334.


\end{thebibliography}
 \end{document}